\documentclass[journal,transmag]{IEEEtran}
\usepackage{graphicx}
\usepackage{float}
\usepackage{amsmath}
\usepackage{amsfonts}
\usepackage{multirow}
\usepackage{subfigure}
\usepackage{color}
\usepackage{comment}
\newcommand {\Define} {\stackrel {\Delta} {=}  }

\begin{document}
%


\title{OTFS $-$ A Mathematical Foundation for 
Communication and Radar Sensing in the Delay-Doppler Domain \\
\thanks{\copyright 2023 IEEE.  Personal use of this material is permitted.  Permission from IEEE must be obtained for all other uses, in any current or future media, including reprinting/republishing this material for advertising or promotional purposes, creating new collective works, for resale or redistribution to servers or lists, or reuse of any copyrighted component of this work in other works.}}



\author{\IEEEauthorblockN{Saif Khan Mohammed\IEEEauthorrefmark{1},
Ronny Hadani\IEEEauthorrefmark{2},
Ananthanarayanan Chockalingam\IEEEauthorrefmark{3}, and 
Robert Calderbank\IEEEauthorrefmark{4},~\IEEEmembership{Fellow,~IEEE}}
\IEEEauthorblockA{\IEEEauthorrefmark{1}Department of Electrical Engineering, Indian Institute of Technology Delhi, India}
\IEEEauthorblockA{\IEEEauthorrefmark{2}Department of Mathematics, University of Texas at Austin, USA}
\IEEEauthorblockA{\IEEEauthorrefmark{3} Department of Electrical and Communication Engineering, Indian Institute of Science, Bangalore, India}
\IEEEauthorblockA{\IEEEauthorrefmark{4}Department of Electrical and Computer Engineering, Duke University, USA}}

%



\IEEEtitleabstractindextext{%
\begin{abstract}
Orthogonal time frequency space (OTFS) is a framework for communication and active sensing that processes signals in the delay-Doppler (DD) domain. This paper explores three key features of the OTFS framework, and explains their value to applications. The first feature is a compact and sparse DD domain parameterization of the wireless channel, where the parameters map directly to physical attributes of the reflectors that comprise the scattering environment, and as a consequence these parameters evolve predictably. The second feature is a novel waveform / modulation technique, matched to the DD channel model, that embeds information symbols in the DD domain. The relation between channel inputs and outputs is localized, non-fading and predictable, even in the presence of significant delay and Doppler spread, and as a consequence the channel can be efficiently acquired and equalized. By avoiding fading, the post equalization SNR remains constant across all information symbols in a packet, so that bit error performance is superior to contemporary multi-carrier waveforms. Further, the OTFS carrier waveform is a localized pulse in the DD domain, making it possible to separate reflectors along both delay and Doppler simultaneously, and to achieve a high-resolution delay-Doppler radar image of the environment. In other words, the DD parameterization provides a common mathematical framework for communication and radar. This is the third feature of the OTFS framework, and it is ideally suited to intelligent transportation systems involving self-driving cars and unmanned ground/aerial vehicles which are self/network controlled. The OTFS waveform is able to support stable and superior performance over a wide range of user speeds. In the emerging 6G systems and standards, it is ideally suited to support mobility-on-demand envisaged in next generation cellular and WiFi systems, as well as high-mobility use cases. Finally, the compactness and predictability of the OTFS input-output relation makes it a natural fit for machine learning and AI algorithms designed for intelligent non-myopic management of control plane resources in future mobile networks.
\end{abstract}

}

\maketitle

\IEEEdisplaynontitleabstractindextext

%
\IEEEpeerreviewmaketitle

%
%
%
%


\section{Introduction}

{
Cellular mobile communication technology evolves as new services and new wireless channels emerge. 
As user demand shifted from simple voice-only services to high speed data, wireless technology shifted from narrowband time-division in 2G systems to frequency-division in 4G and 5G systems \cite{WC3a}-\cite{WC3d}. To some degree, it is innovation at the physical layer that distinguishes one generation from the next. 
Mobile communications systems serving billions of people all over the world 
have been made possible by theoretical advances in coding \cite{inno1},\cite{inno2}, in methods that provide spatial diversity \cite{inno3}-\cite{inno8}, and in methods of user cooperation \cite{inno9},\cite{inno10}. 
This flow of ideas has transformed the teaching of wireless communications 
\cite{WC1}-\cite{WC4}.
We expect that innovation at the physical layer will continue as carrier frequencies increase (mmWave and THz), and new high-mobility use cases emerge (bullet trains, autonomous vehicles, airplanes). However, perhaps the most fundamental change in mobile communications is the relative importance of the physical layer, the network layer and the services layer. As the IP revolution transformed the wireless world, cell phones migrated to smartphones which are complex software platforms. Demand for capacity could not be met by circuit switched networks engineered to provide worst case coverage at the cell boundary, and the physical layer changed in response \cite{WC3}. We emphasize predictability in this article because a physical layer that is predictable simplifies network management and the provision of services.

Enhancements to mobile communication services across wireless generations have been enabled by judicious allocation of time and frequency resources. To date, these enhancements have been achieved primarily within the framework of time domain modulation (TDM) or frequency domain modulation (FDM). In TDM based schemes, information is carried by a narrow time domain (TD) pulse, and an information packet is a superposition of such TD pulses. In TDM, the signal is localized in the TD but not in the frequency domain (FD), and as a consequence, the interaction between a TD pulse and the environment leads to time selective fading. Likewise, in FDM based schemes, information is carried by a narrow FD pulse, which is essentially a sinusoid in the time domain. The FDM signal is localized in the FD but not in the TD, and this lack of localization in the TD causes frequency selective fading. The lack of time frequency (TF) localization in TDM/FDM signals also makes the relation between channel inputs and channel outputs less predictable. Fading and lack of predictability degrade the performance of TDM/FDM on doubly-spread channels, that is, channels that are selective in both time and frequency \cite{WC4}. As carrier frequencies increase and high-speed use cases emerge, we encounter doubly-spread channels that are more extreme. TDM/FDM have served well in wireless generations to date, but better TF localization may be required in the wireless generations to come.

TF localization is particularly important to emerging intelligent transportation systems, where accurate and high-resolution radar imaging can identify potential hazards in fast-moving environments and enable responses that enhance road safety \cite{RS1}. Waveforms with better TF localization enable more accurate location of objects like pedestrians and bicycles which have small radar cross sections. Looking back to 1953, only 5 years after Claude Shannon created information theory, Philip Woodward described how to think of radar in information theoretic terms \cite{RS2}. He suggested that we view the radar scene as an unknown operator parametrized by delay and Doppler, and that we view radar waveforms as questions that we ask the operator. Woodward proposed to define a good question in terms of lack of ambiguity in the answer, and he sought questions with good localization in delay and Doppler. By identifying a single waveform, good for both channel estimation and communication, we seek to decrease system complexity, reduce electromagnetic interference, and mitigate spectrum related concerns \cite{RS3}. 

We describe wireless channels in terms of delay and Doppler operators, and we describe how to use certain geometric modes of these operators for communication and sensing. These special modes constitute the orthogonal time frequency space (OTFS) waveforms, introduced in \cite{RH1} (see also \cite{RH2,RH3}). When viewed in the time domain, an OTFS waveform is a pulse train modulated by a tone, a pattern that we refer to as a pulsone and we provide a comprehensive review of its mathematical foundation. The heart of this foundation is the Zak transform, which converts the TD signal into a quasi-periodic function in delay and Doppler. We describe how OTFS waveforms are pulses in the DD domain, that are engineered to mirror the dynamics of the wireless channel. We derive their TF localization properties from first principles, explaining why the relation between channel inputs and channel outputs is predictable, and why fading is eliminated. We also explain how this approach to TF localization provides a geometric interpretation of the Nyquist rate. In the context of radar, we explain why Woodward might have thought of OTFS waveforms as good questions. In the context of communications, we explain how OTFS waveforms give rise to a modulation scheme that multiplexes information in the DD domain. The wireless channel is determined by a small number of dominant reflectors, which means that in the DD domain, it admits a sparse representation. Moreover, this representation changes only at the speed of the underlying physics of the reflectors which renders it ``effectively" stationary. In this context, we explain how better TF localization of the carrier waveform takes advantage of this DD domain stationary representation of the channel, translating to better performance under doubly-spread channels.

This is the first of two papers, providing the mathematical foundation for signal processing in the delay-Doppler domain.} {It prepares the ground for the second paper which
studies the performance of OTFS in comparison to other modulation schemes such as TDM and FDM, and explores the utility of the OTFS waveform for radar sensing.}

\section{The Delay-Doppler Domain}
\label{secWirelessChannel}
\begin{figure*}[!h]
\centering
\includegraphics[width=16cm, height=8.0cm]{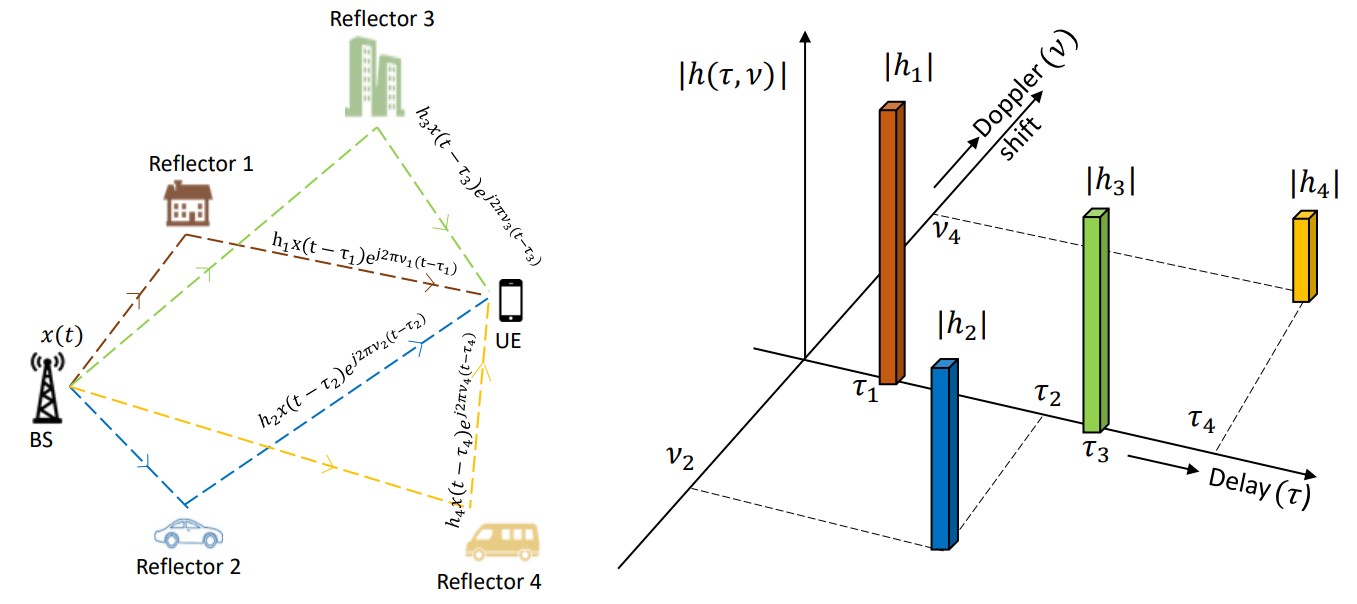}
\caption{{
The delay-Doppler spreading function $h(\tau,\nu)$ of a doubly-spread wireless channel. Four dominant paths between the base station (BS) and the user equipment (UE) result in a function $h(\tau,\nu)$ comprising impulses at $(\tau,\nu) = (\tau_i,\nu_i)$ where $\tau_i$ is the $i$th path delay and $\nu_i$ is the $i$th path Doppler shift.
Note that the first and third reflectors do not introduce Doppler shifts since they are stationary.  Also, the Doppler shift induced by the second and the fourth reflector have opposite
polarity as these mobile reflectors are travelling in opposite directions.
}}
\label{fig_1}
\end{figure*}
{A typical} wireless channel between a transmitter and a receiver is determined by a relatively small number of dominant propagation paths. Fig.~\ref{fig_1} shows a time-domain (TD) signal $x(t)$ transmitted by a base station (BS) and received at some user equipment (UE) through four propagation paths. Each path is characterized by the \emph{path delay}, which is the time taken by the signal to propagate along the path, and by the \emph{path Doppler shift}, which is the frequency shift induced
by the relative motion of transmitter, reflector, and receiver.
In Fig.~\ref{fig_1}, we assume the BS and UE are stationary so that only reflector motion determines path Doppler shift.
 The signal copy received at UE along the $i$th path is given by $h_i x(t - \tau_i) \, e^{j 2 \pi \nu_i (t - \tau_i)}$, where
$h_i$ is the complex channel path gain, $\tau_i$ seconds is the path delay, and
$\nu_i$ Hz is the path Doppler shift.

We refer to wireless channels as \emph{doubly-spread}, characterized by the delay, Doppler shift, and complex channel gain of each channel path. The action of the wireless channel on the transmitted signal $x(t)$ is specified by the
\emph{delay-Doppler (DD) spreading function}
$h(\tau,\nu)$, where $\tau \in {\mathbb R}$
and $\nu \in {\mathbb R}$ are the delay and Doppler variables, respectively. In Fig.~\ref{fig_1}
\begin{eqnarray}
\label{eqn1}
h(\tau, \nu) & = & \sum\limits_{i=1}^4 \, h_i \, \delta(\tau - \tau_i) \, \delta(\nu - \nu_i),
\end{eqnarray}
where $\delta(\cdot)$ denotes the Dirac-delta impulse function.
The noise-free TD signal received at the UE is given by \cite{Bello63}
\begin{eqnarray}
\label{eqn2}
y(t) & = & \int\int h(\tau,\nu) \, x(t - \tau) \, e^{j 2 \pi \nu (t - \tau)} \, d\tau \, d\nu.
\end{eqnarray}
In Fig.~\ref{fig_1},
\begin{eqnarray}
\label{eqn3}
y(t) & = & \sum\limits_{i=1}^4 h_i \, x(t - \tau_i) \, e^{j 2 \pi \nu_i (t - \tau_i)},
\end{eqnarray}
which is the sum of the signal copies received along the four paths.

\section{Time and Frequency Domain Modulation}
\label{sec2}
\begin{figure*}[!h]
\centering
\includegraphics[width=14cm, height=8.0cm]{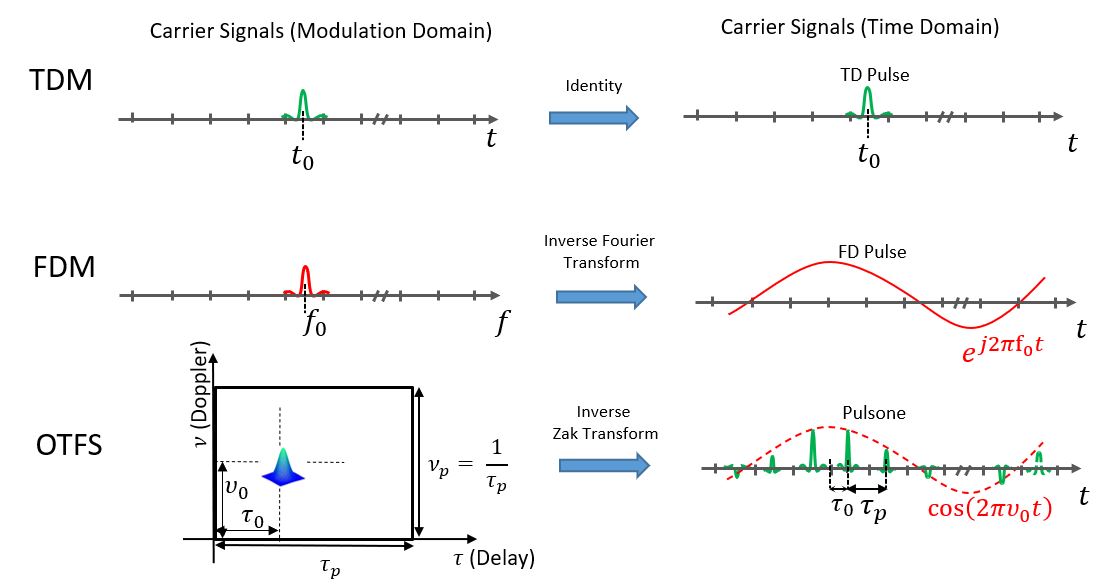}
\caption{Information carriers for TDM/FDM/OTFS in modulation domain and time domain. Traditional TDM and FDM carriers are narrow pulses in TD and FD but spread in FD and TD respectively, manifesting the fundamental obstruction for TF localization. In contrast, the OTFS carrier is a quasi-periodic pulse in the DD domain, viewed as ``effectively" localized jointly in time and frequency.}
\label{fig_2}
\end{figure*}
In time domain modulation (\emph{TDM}), information is carried by a \emph{narrow TD pulse}, and an information packet is the superposition of distinct TD pulses, each modulated by an information symbol. 
In frequency domain modulation (\emph{FDM}), information is carried by a \emph{narrow FD pulse}, which is essentially a sinusoid in the time domain. An information packet is a TD signal, which is the inverse Fourier transform of a superposition of distinct FD pulses, each modulated by an information symbol.

The information carrier in TDM is localized in the TD but not in the FD. Similarly, the information carrier in FDM is localized in the FD but not in the TD (the sinusoid $e^{j 2 \pi f_0 t}$ shown in Fig.~\ref{fig_2} is spread in time, but localized at $f = f_0$ in frequency). {As will be explained in the sequel, the implication of this lack of localization is that the TDM/FDM input-output relation under doubly-spread channel witnesses fading and non-predictability.
The fading attribute implies a degradation in the BER performance while the non-predictability attribute implies
frequent acquisition of the effective channel response
in order to maintain good performance (see Section \ref{secTDMFDMIOrelation} and Section \ref{secFDMIOrelation}).}
{The Heisenberg uncertainty principle implies that a signal cannot be simultaneously localized jointly in time and frequency. However, as will be explained in the sequel, this obstruction can be ``effectively'' eliminated as long as a certain quasi-periodic condition is maintained (see Section \ref{secDDCarrier}). 
In OTFS, information is carried by a \emph{DD domain} quasi-periodic pulse. The translation from a DD signal into a TD signal is carried by the \emph{inverse Zak transform} (see Section \ref{secDDCarrier}), just as the translation from an FD signal into a TD signal is carried by the inverse Fourier transform.
The bottom part of Fig.~\ref{fig_2} depicts the TD realization of a DD domain pulse. We refer to this TD signal as a \emph{pulsone}, since it is essentially comprised of a pulse train modulated by a frequency tone (see also \cite{RH3} and \cite{SKM1})\footnote{\footnotesize{{As we shall see later in Section \ref{secDDCarrier}, the ``effective" localization of a pulsone (of approximate bandwidth $B$ and time duration $T$) is not exact and that the corresponding DD domain pulse has most of its energy spread in a narrow region having width $B^{-1}$ and $T^{-1}$ along the delay and Doppler axis, respectively.} }}.
The main beneficial property of the pulsone is that its interaction with a doubly-spread channel is both non-fading and predictable, which translates to superior BER performance and more efficient channel acquisition (see Section \ref{secOTFSIOrelation}).}


\section{Delay-Doppler Modulation}
\label{secDDCarrier}
\begin{figure}[!h]
\centering
\includegraphics[width=7.5cm, height=5.5cm]{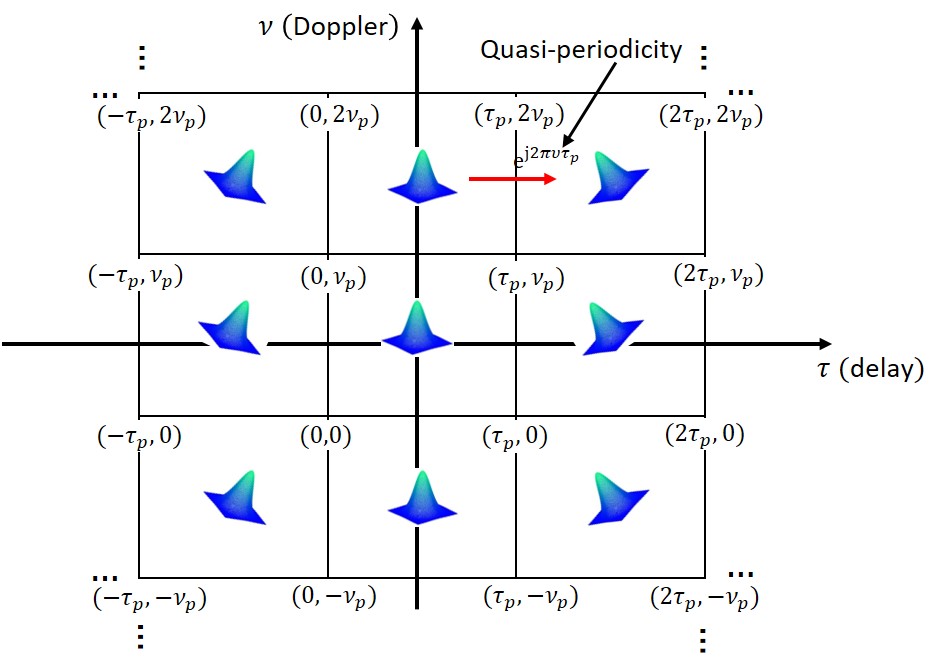}
\caption{A DD domain pulse is localized only within the fundamental DD domain period ${\mathcal D}_0$, as it repeats infinitely many times in a quasi-periodic fashion.}
\label{fig_3}
\end{figure}
{As mentioned in the previous section, in OTFS, information is carried over a DD domain pulse. In this section, we give the formal definition of a DD domain pulse and study its local structure. In addition, we illustrate the impact of various DD domain pulse parameters on its TD and FD realizations. Finally, we derive the global properties of the OTFS modulation from the local properties of a DD domain pulse. The definition of a DD domain pulse depends on a choice of two
periods: the \emph{delay period}, denoted by $\tau_p \in {\mathbb R}_{\geq 0}$ and the \emph{Doppler period}, denoted by $\nu_p \in {\mathbb R}_{\geq 0}$. The two periods should be reciprocal, that is, $\nu_p = 1/\tau_p$. For any given choice of the periods, there is an associated transform, called the time Zak transform, denoted by ${\mathcal Z}_t$, that establishes a unitary equivalence between TD signals and a sub-class of quasi-periodic DD domain signals.\footnote{\footnotesize{{The definition of the time Zak transform depends on the values of the periods. For clarity, we omit the periods from the notation, assuming their values to be clear from the context.}}}} {The time Zak transform of a TD signal $x(t)$ is given by} 
\begin{eqnarray}
\label{eqn7}
x_{_{\mbox{\footnotesize{dd}}}}(\tau,\nu) & = & {\mathcal Z}_t( x(t)) \nonumber \\
& \Define & \sqrt{\tau_p} \sum\limits_{k=-\infty}^{\infty}  x(\tau + k \tau_p) \, e^{-j 2 \pi \nu k \tau_p}.
\end{eqnarray}
{Observe from (\ref{eqn7}) that,
for any $n,m \in {\mathbb Z}$,
$x_{_{\mbox{\footnotesize{dd}}}}(\tau,\nu)$ satisfies} 

\begin{figure*}[!h]
\centering
\includegraphics[width=14cm, height=8.0cm]{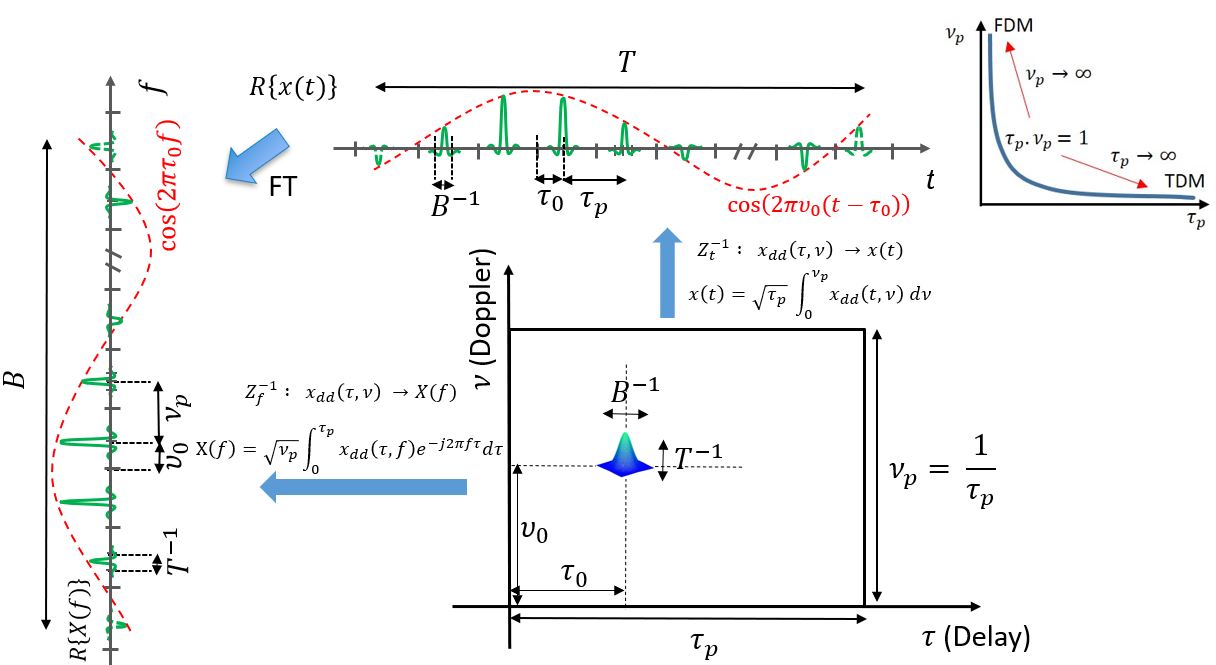}
\caption{{A DD domain pulse and its TD/FD realizations referred to as TD/FD pulsone. The TD pulsone comprises of a finite duration pulse train modulated by a TD tone. The FD pulsone comprises of a finite bandwidth pulse train modulated by a FD tone. The location of the pulses in the TD/FD pulse train and the frequency of the modulated TD/FD tone is determined by the location of the DD domain pulse $(\tau_0,\nu_0)$. The time duration and bandwidth of a pulsone are inversely proportional to the characteristic width of the DD domain pulse along the Doppler axis and the delay axis, respectively. The number of non-overlapping DD pulses, each spread over an area $B^{-1}T^{-1}$, inside the fundamental period ${\mathcal D}_0$ (which has unit area) is equal to the time-bandwidth product $BT$ and the corresponding pulsones are orthogonal to one another, rendering OTFS an orthogonal modulation that achieves the Nyquist rate.  As $\nu_p \rightarrow \infty$, the FD pulsone approaches a single FD pulse which is the FDM carrier. Similarly, as $\tau_p \rightarrow \infty$, the TD pulsone approaches a single TD pulse which is the TDM carrier. OTFS is therefore a family of modulations parameterized by $\tau_p$ that interpolates between TDM and FDM.}}
\label{fig_4}
\end{figure*}

{\vspace{-4mm}
\small
\begin{eqnarray}
\label{eqn4}
x_{_{\mbox{\footnotesize{dd}}}}(\tau + n \tau_p, \nu + m \nu_p) &  & \nonumber \\
& \hspace{-24mm} = & \hspace{-13mm}
\sqrt{\tau_p} \hspace{-1mm} \sum\limits_{k=-\infty}^{\infty}  \hspace{-1mm} x(\tau + (k + n) \tau_p) \, e^{-j 2 \pi (\nu + m \nu_p) k \tau_p} \nonumber \\
& \hspace{-24mm} = & \hspace{-13mm} \sqrt{\tau_p} \sum\limits_{k^\prime=-\infty}^{\infty}  x(\tau + k^\prime \tau_p) \, e^{-j 2 \pi \nu k^\prime \tau_p} e^{j 2 \pi n \nu \tau_p} \nonumber \\
& \hspace{-24mm} = & \hspace{-13mm}  e^{j 2 \pi n \nu \tau_p} \, x_{_{\mbox{\footnotesize{dd}}}}(\tau, \nu) \,\,,\,\, n,m \in {\mathbb Z}.
\end{eqnarray}\normalsize}
\hspace{-4mm}
{The condition in (\ref{eqn4}) is referred to as the quasi-periodicity condition.
Fig.~\ref{fig_2} depicts a quasi-periodic pulse which is localized at the point $(\tau,\nu) = (\tau_0,\nu_0)$
within the rectangular region}
\begin{eqnarray}
{\mathcal D}_0 & \Define & \{ (\tau,\nu) \, {\Big \vert}  \,  0 \leq \tau < \tau_p, \, 0 \leq \nu < \nu_p \}.
\end{eqnarray}
{We refer to ${\mathcal D}_0$ as the \emph{fundamental period} of the DD domain.
Due to its quasi-periodicity (see (\ref{eqn4})), the pulse is present at all integer translates
$(\tau,\nu) = (\tau_0 + n \tau_p, \nu_0 + m \nu_p)$, where $m,n \in {\mathbb Z}$, as shown
in Fig.~\ref{fig_3}.}
Note that the phase of the pulse
changes when the pulse location shifts by an integer multiple of $\tau_p$ along the delay axis, but there is no change in phase when the pulse location shifts by an integer multiple of $\nu_p$ along the Doppler axis. In summary, the DD domain pulse comprises of a configuration of infinitely many pulses which repeat at integer multiples of $\tau_p$ along the delay axis, and at integer multiples of $\nu_p = 1/\tau_p$ along the Doppler axis.
{The center part of Fig.~\ref{fig_4} depicts a DD domain pulse $x_{_{\mbox{\footnotesize{dd}}}}(\tau , \nu )$ located at $(\tau,\nu) = (\tau_0 , \nu_0 )$
within the fundamental period ${\mathcal D}_0$. Along the delay axis, it is spread over a
characteristic length $1/B < \tau_p$, and along the Doppler axis, it is spread over a characteristic length $1/T < \nu_p$. Since the DD domain pulse is quasi-periodic, it repeats infinitely many times
as shown in Fig.~\ref{fig_3}.}
{We first analyze the structure of the TD realization,
which is obtained by applying the inverse time Zak transform \cite{Zak67, Janssen88}, i.e.,}

{\small
\vspace{-4mm}
\begin{eqnarray}
\label{eqn6}
x(t) & = & {\mathcal Z}_t^{-1}\left( x_{_{\mbox{\footnotesize{dd}}}}(\tau,\nu) \right)  \, \Define \, \sqrt{\tau_p} \int_{0}^{\nu_p} x_{_{\mbox{\footnotesize{dd}}}}(t,\nu) \, d\nu.
\end{eqnarray}\normalsize}
\hspace{-4mm}
The top part of Fig.~\ref{fig_4} shows that $x(t)$ is a pulse train of finite duration $T$. Each pulse in the train is spread over a time duration $1/B$, and consecutive pulses are separated by the delay period $\tau_p$. The pulses are located at time instances $t = n \tau_p + \tau_0$, $n \in {\mathbb Z}$, where $\tau_0$ is the delay coordinate of the underlying DD domain pulse. The pulse train is modulated by a sinusoid of frequency $\nu_0$, where $\nu_0$ is the Doppler coordinate of the underlying DD domain pulse.

{Next, we analyze the structure of the FD realization, which is obtained by applying the inverse frequency Zak transform ${\mathcal Z}_f^{-1}$ \cite{Zak67, Janssen88}, i.e.,}

{\small
\vspace{-4mm}
\begin{eqnarray}
\label{eqn8}
\hspace{-2mm}
X(f) & \hspace{-2mm} = & \hspace{-2mm} {\mathcal Z}_f^{-1}( x_{_{\mbox{\footnotesize{dd}}}}(\tau,\nu))  \, \Define \, \sqrt{\nu_p} \int_{0}^{\tau_p} x_{_{\mbox{\footnotesize{dd}}}}(\tau,f) \, e^{-j 2 \pi f \tau} d\tau.
\end{eqnarray}\normalsize}
\hspace{-3mm}
The left part of Fig.~\ref{fig_4} shows that $X(f)$ is also a pulsone, comprising of a pulse train which extends over a bandwidth $B$. Each pulse in the train is spread over a frequency interval $1/T$. Consecutive pulses are separated by the Doppler period $\nu_p$. The pulses are located at frequencies $f = m \nu_p + \nu_0$, $m \in {\mathbb Z}$, where $\nu_0$ is the Doppler coordinate of the underlying DD domain pulse.
The pulse train is modulated by a FD sinusoid $e^{-j 2 \pi \tau_0 f }$ where $\tau_0$ is the delay coordinate of the underlying DD domain pulse.  

\begin{figure*}[!h]
\centering
\includegraphics[width=15cm, height=7.0cm]{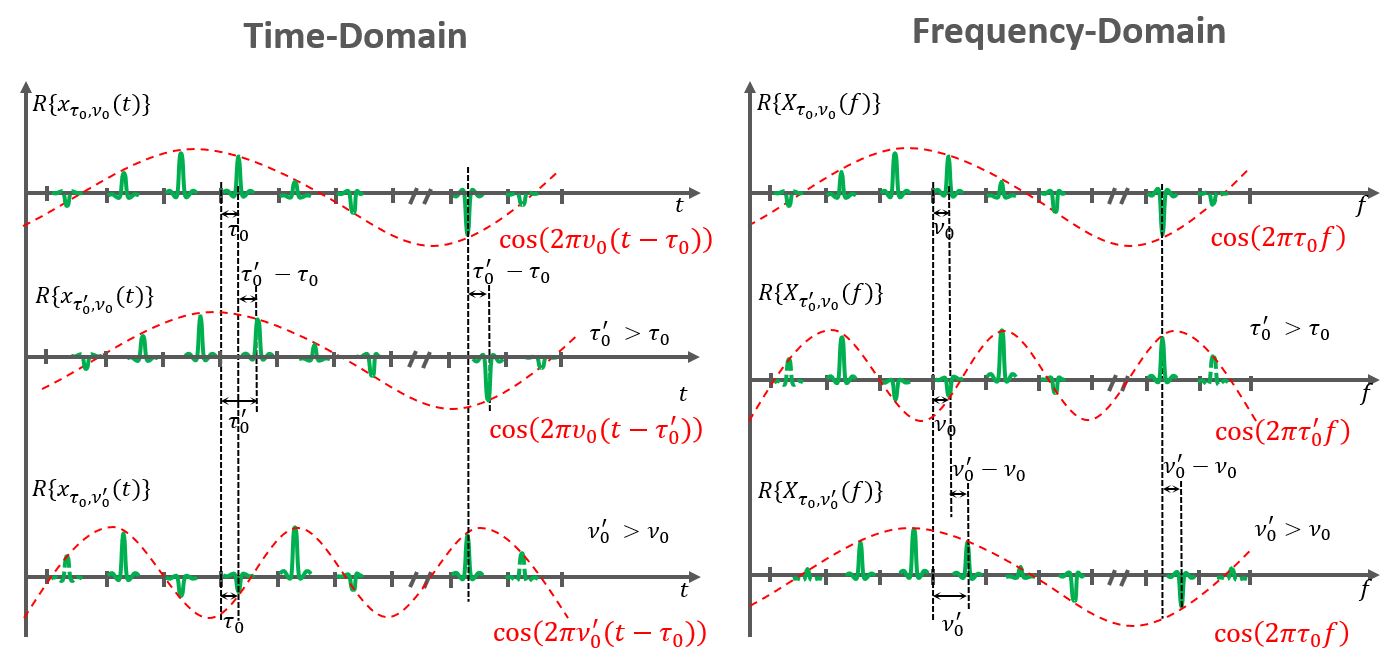}
\caption{{Impact of DD domain pulse location on the TD/FD pulsone characteristics. The figure consists of six plots organized into three rows and two columns (one column for TD and the other for FD). The TD plot in the second row of the figure shows that, a shift in the delay axis location of a DD domain pulse from $\tau_0$ to $\tau_0'$ translates to a time displacement of the TD pulsone by $(\tau_0' - \tau_0)$ seconds. The effect of this shift on the corresponding FD pulsone (shown on the right in the same row) is that the modulating FD tone changes from $e^{-j 2 \pi \tau_0 f}$ to $e^{-j 2 \pi \tau_0' f}$ (only the real part of the TD/FD pulsones is plotted). The FD plot in the third row of the figure shows that, a shift in the Doppler axis location of a DD domain pulse from $\nu_0$ to $\nu_0'$ translates to a frequency displacement of the FD pulsone by $(\nu_0' - \nu_0)$ Hz. The effect of this shift on the corresponding TD pulsone (shown on the left in the same row) is that the modulating TD tone changes from $e^{j 2 \pi \nu_0 (t - \tau_0)}$ to $e^{j 2 \pi \nu_0' (t - \tau_0)}.$}}
\label{fig_5a}
\end{figure*}

\begin{figure*}[!h]
\centering
\includegraphics[width=14cm, height=6.0cm]{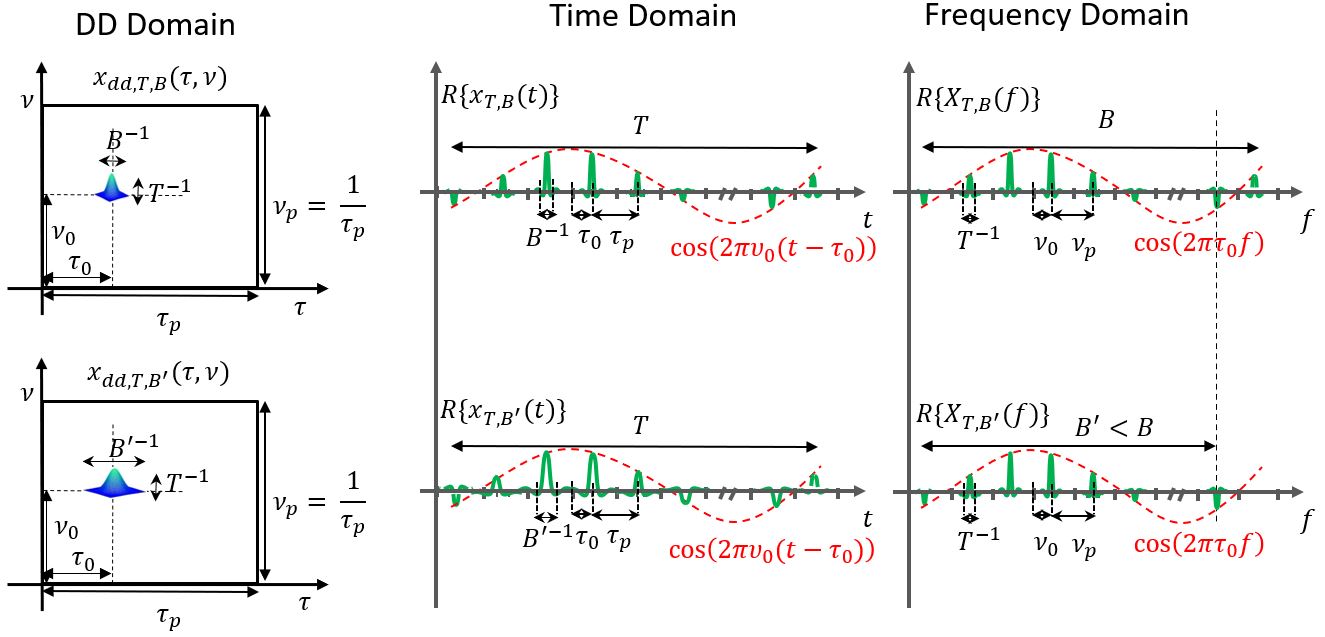}
\caption{{Impact of DD domain pulse width along delay axis on the TD/FD pulsone characteristic. Increasing the width of the DD domain pulse along the delay axis from $\frac{1}{B}$ to $\frac{1}{B'}$ translates to a reduction of the FD pulsone bandwidth from $B$ to $B'$ and an increase in the width of each TD pulse in the corresponding TD pulsone from $\frac{1}{B}$ to $\frac{1}{B'}$.}}
\label{fig_5b}
\end{figure*}

\begin{figure*}[!h]
\centering
\includegraphics[width=14cm, height=6.0cm]{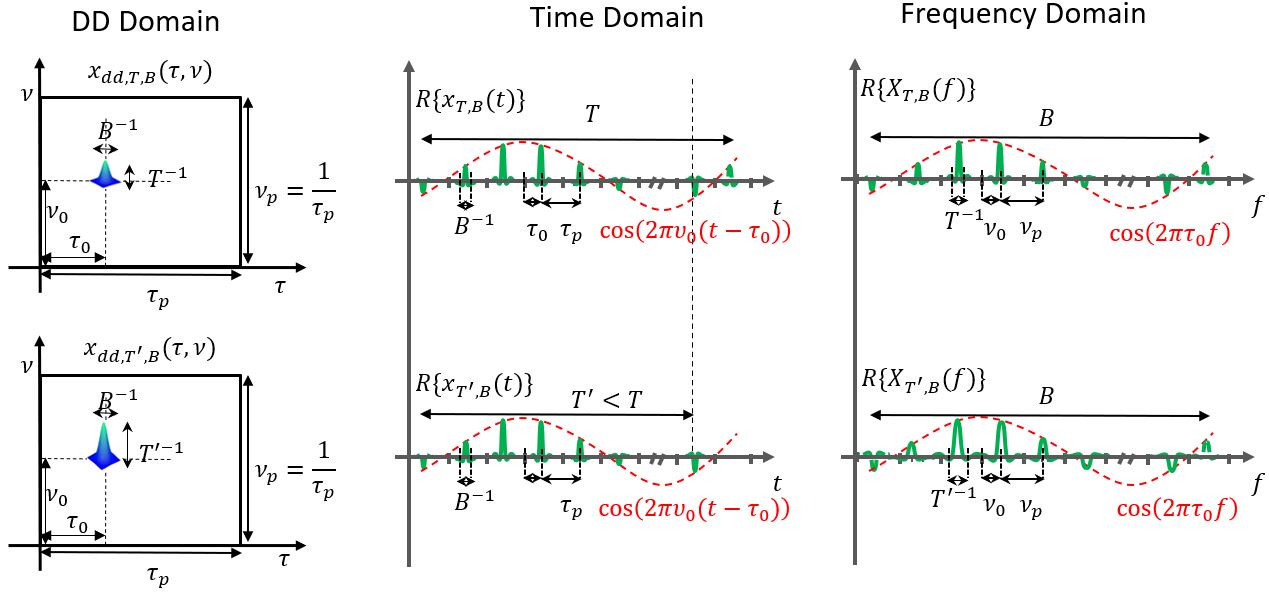}
\caption{{Impact of DD domain pulse width along the Doppler axis on TD/FD pulsone characteristic. Increasing the width of the DD domain pulse along the Doppler axis from $\frac{1}{T}$ to $\frac{1}{T'} $ translates to a reduction of the TD pulsone time duration from $T$ to $T'$ and an increase in the width of each FD pulse in the corresponding FD pulsone from $\frac{1}{T}$ to $\frac{1}{T'}$.}}
\label{fig_5c}
\end{figure*}

Fig.~\ref{fig_5a} depicts the effect of shifting the location of the DD domain pulse on the structure of the corresponding TD and FD pulsones. In a nutshell, a shift of the DD domain pulse along the delay axis translates to a time displacement of the TD pulsone, and a shift of the DD domain pulse along the Doppler axis translates to a frequency displacement of the FD pulsone.
Fig.~\ref{fig_5b} and Fig.~\ref{fig_5c} depict the effect of increasing the width of the DD domain pulse on the structure of the corresponding TD and FD pulsones. In a nutshell, Fig.~\ref{fig_5b} shows that an increase of the width of the DD domain pulse along the delay axis translates to a reduction of the FD pulsone bandwidth, and Fig.~\ref{fig_5c} shows that an increase of the width of the DD domain pulse along the Doppler axis translates to a reduction of the TD pulsone time duration.

We now derive global properties of the OTFS modulation
from the local properties described above.

\underline{Orthogonality of pulsones:}
{From Fig.~\ref{fig_4} we see that
a DD domain pulse is localized within a rectangular DD domain region which
is $1/B$ seconds wide along the delay axis and $1/T$ Hz wide along the Doppler axis.
Geometrically, this implies almost no overlap between two DD pulses whose delay domain locations differ by $1/B$ or whose Doppler domain locations differ by $1/T$, i.e., such pulses
are almost orthogonal.}

{The same observation can be made by separately observing the TD and FD realizations of a DD pulse.
We know from Fig.~\ref{fig_5a}, that shifting the location of the DD domain pulse by $1/B$ along the delay axis induces a time displacement of the TD pulsone by $1/B$ seconds. From Fig.~\ref{fig_4}, we also know that each pulse in the TD pulsone has time duration $1/B$ and that consecutive pulses are separated by $\tau_p$. When a DD domain pulse is shifted by $1/B$ along the delay axis the corresponding TD pulsones do not overlap.
Similarly, when a DD domain pulse is shifted by $1/T$ Hz along the Doppler axis, the corresponding FD pulsones do not overlap. Delay shifts by integer multiples of $1/B$ and Doppler shifts by integer multiples of $1/T$ lead to TD/FD
pulsones that are almost orthogonal. Here, the impact of delay and Doppler shifts on the orthogonality between pulsones is understood \emph{separately} through
their TD and FD realizations, respectively. The geometric interpretation of
orthogonality in the previous paragraph is however much \emph{simpler} since we
view the pulsone as a pulse in the DD domain.}

\underline{Optimality as time- and bandlimited signals:}
Slepian, Landau and Pollack \cite{Slepian62, Landau62}
introduced the family of prolate spheroidal waveforms
to measure the space of essentially time- and bandlimited
signals. For signals limited approximately to a time duration of $T$ seconds and bandwidth $B$ Hz, they showed that the number of orthogonal carrier waveforms is approximately
equal to the time-bandwidth product $B T$. We use pulsones
with time-duration $T$ and bandwidth $B$ to span this same space. Recall that pulsones are almost orthogonal if the location of the corresponding DD domain pulses differ by integer multiples
of $1/B$ along the delay axis and by integer multiples of
$1/T$ along the Doppler axis. The number of approximately
orthogonal carrier waveforms is essentially $\frac{\tau_p \nu_p}{\frac{1}{B} \frac{1}{T}} = BT$, which is the time-bandwidth product. 

\underline{TDM as a limiting case:}
In the TD pulsone shown in Fig.~\ref{fig_4}, as $\tau_p \rightarrow \infty$,
the TD pulses located at $t = \tau_0 + n \tau_p, n \in {\mathbb Z}, n \ne 0,$
move towards $\pm \infty$ and only the TD pulse at $t = \tau_0$ remains, i.e., 
the TD pulsone approaches a single TD pulse which is the TDM carrier. In other words, as the Doppler domain collapses, the OTFS carrier tends to the TDM carrier.

\underline{FDM as a limiting case:}
In the FD pulsone, as $\nu_p \rightarrow \infty$ (i.e., $\tau_p = (1/\nu_p) \rightarrow 0$), the pulses located at $f = \nu_0 + m \nu_p, m \in {\mathbb Z}, m \ne 0,$ move towards $\pm \infty$ and only the FD pulse at $f=\nu_0$ remains, i.e., the FD pulsone approaches a single FD pulse which is the FDM carrier.
In other words,
as the delay domain collapses, the OTFS carrier tends to the FDM carrier.

\emph{OTFS is therefore a family of modulations parameterized by $\tau_p$ that interpolates between TDM and FDM} (see the hyperbola $\tau_p \, \nu_p = 1$ in the top right corner of Fig.~\ref{fig_4}).\\

\underline{TD pulsones encode wireless channel dynamics:}
TD pulsones are engineered to mirror the dynamics of the wireless channel. The effect of channel path delay on a DD domain pulse is to simply shift the pulse along the delay axis by an amount equal to the path delay. The effect of a channel path Doppler shift on a DD domain pulse is to simply
shift the pulse along the Doppler axis by an amount equal to the path Doppler shift.

\begin{figure}[!h]
\centering
\includegraphics[width=9cm, height=6.0cm]{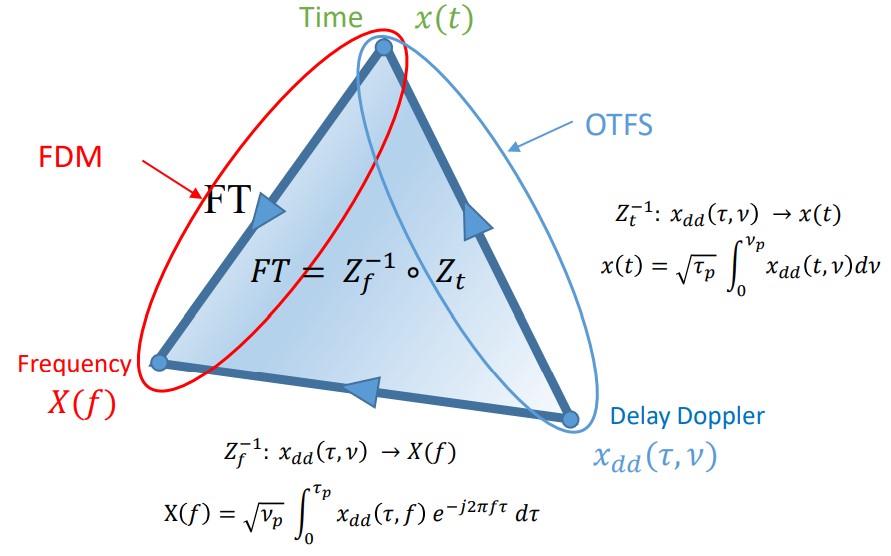}
\vspace{-1mm}
\caption{Three different basic signal realizations, TD, FD and DD domains. Signal representations in these
domains are related through transforms. The well known Fourier transform is in fact a composition of the Zak transform ${\mathcal Z}_t$ and the inverse Zak transform ${\mathcal Z}_f^{-1}$ from DD domain to FD.}
\label{fig_6}
\end{figure}

\underline{The Fourier transform as a composition:}
Fig.~\ref{fig_4} illustrates that we can map a TD signal to
its FD realization by first applying the Zak transform ${\mathcal Z}_t$ from the TD to the DD domain, then applying
the inverse frequency Zak transform ${\mathcal Z}_f^{-1}$ from the
DD domain to the FD. In other words, the Fourier transform
is the composition of ${\mathcal Z}_t$ and ${\mathcal Z}_f^{-1}$. Fig.~\ref{fig_6} represents the three signal representations (TD, FD and DD domain) as the three vertices of a triangle, and labels edges between vertices by the transforms between signal representations.


\section{Interaction of channel paths and carrier waveforms}
\label{secChannelInteraction}
In this section, we use the example shown in Fig.~\ref{fig_7}, to illustrate how a doubly-spread wireless channel interacts with the carrier waveforms for TDM, FDM and OTFS. {The doubly-spread wireless channel considered in Fig.~\ref{fig_7} comprises of four paths. The first and the third paths are due to reflection from stationary building and therefore these paths do not result in any Doppler shift. The second and the fourth paths are due to reflection from moving vehicles and these therefore induce Doppler shift. To highlight the
phenomenon of fading and non-predictability of channel interaction in TDM, we consider the first and the second path to have the same delay so that the TD pulses received along these two paths superimpose in a time-varying manner (due to the Doppler shift of the second path). Similarly, choosing the first and the third path to have the same Doppler shift (i.e., zero) but different delay, highlights the phenomenon of fading and non-predictability of channel interaction in FDM. However,
since any two paths differ in either delay or Doppler shift, in OTFS there is no superposition of DD domain pulses received along distinct paths. Due to this reason, for suitable values of $(\tau_p,\nu_p)$ the interaction of a DD pulse with the channel is non-fading and predictable. The exact values of the path gain, delay and Doppler shifts is mentioned inside the parameter box in Fig.~\ref{fig_7}.}

\subsection{Interaction with a TD pulse}
\label{secChannelInteractionTD}
\begin{figure*}[!h]
\centering
\includegraphics[width=16cm, height=8.0cm]{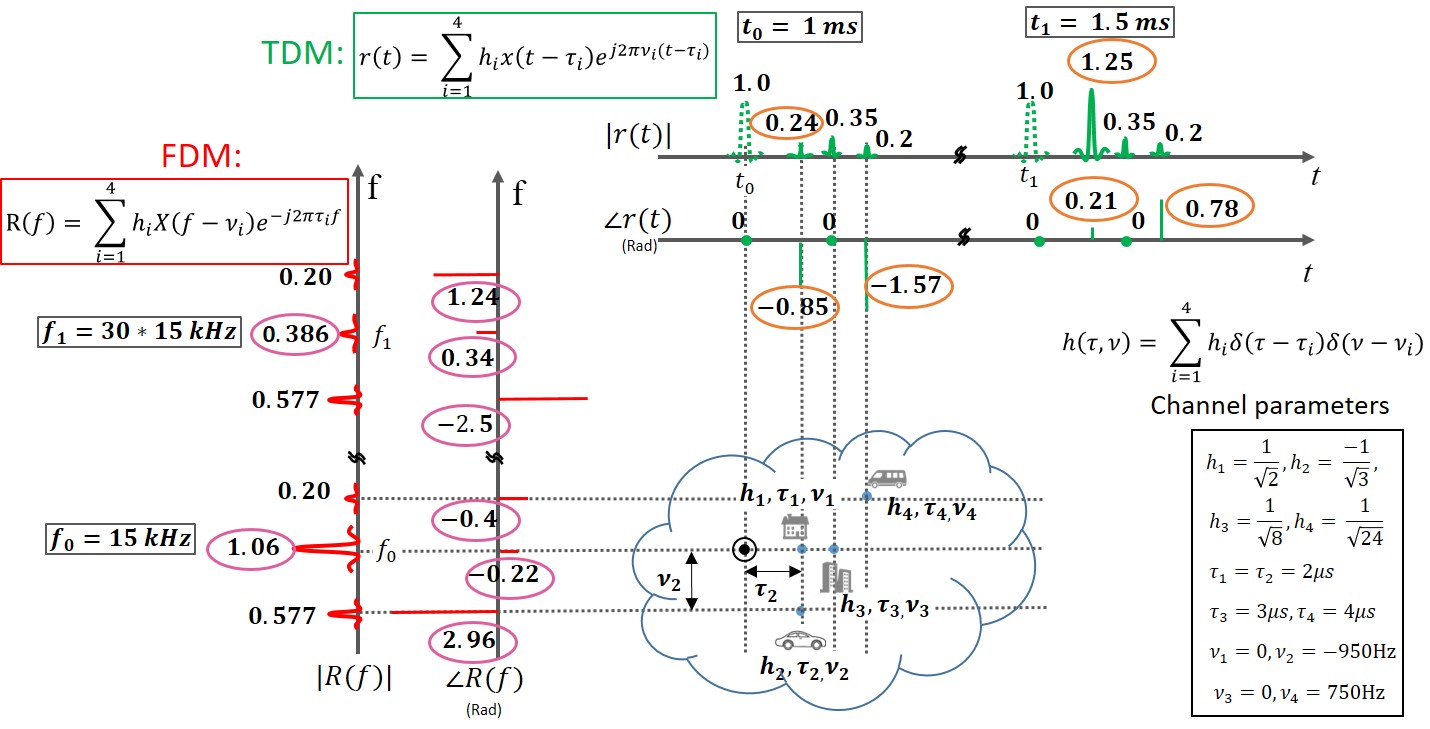}
\caption{Interaction of a doubly-spread channel with TDM/FDM carrier waveform. Two TD pulses are transmitted at $t=t_0$ and $t= t_1$. Due to the first and the second path (which have the same propagation delay), pulses are received at $t = t_0 + 2 \mu s$ and $t = t_1 + 2 \mu s$. Due to the different Doppler shifts of these two paths, the magnitude and phase of these received pulses is dependent on the time instances $t_0$ and $t_1$ in a non-simple manner. Therefore, the magnitude and phase
of the pulse received at $t= t_1 + 2 \mu s$ \emph{cannot} be simply predicted given the knowledge of the pulse received at $t = t_0 + 2 \mu s$. As the magnitude and phase of these two received pulses are different, this interaction is \emph{non-predictable, fading and non-stationary}. Similarly,
two FD pulses are transmitted at $f=f_0$ and $f=f_1$. Due to the first and the third path which induce the same zero Doppler shift and have different path delays, the magnitude and phase of the received FD pulses along these two paths depends on the FD location of the transmitted FD pulse. This dependence is non-simple and therefore the magnitude and phase of the FD pulse received at $f=f_1$ cannot be predicted based on the knowledge of the magnitude and phase of the FD pulse received at $f=f_0$. The dependence on the FD pulse location is due to the non-zero path delays of these two paths, which results
in a \emph{non-predictable, fading and non-stationary} interaction between the channel and the FD pulse. The \emph{non-predictable, fading and non-stationary} interaction between the TD/FD pulse and the channel is not because of the fixed underlying channel (i.e., path gain, delays and Doppler shifts), but is because these pulses are not simultaneously localized in TD and FD.} 
\label{fig_7}
\end{figure*}
Fig.~\ref{fig_7} illustrates the TD response to a transmitted signal
\begin{eqnarray}
\label{eqn9}
x(t) & = & \delta(t - t_0) \, + \, \delta(t - t_1).
\end{eqnarray}
comprising two narrow TD pulses\footnote{\footnotesize{We represent the narrow pulses by Dirac-delta impulses since the width of a pulse is much smaller than the channel path delays and the inverse of the maximum Doppler shift.}}
transmitted at $t=t_0 = 1$ ms and
$t=t_1 = 1.5$ ms. We consider a stationary doubly-spread channel
comprising four channel paths with
parameters $h_i, \tau_i, \nu_i$ listed in the channel parameter box.
We let $h(t; t_i), i=0,1$ denote the TD impulse response of the channel to a TD pulse transmitted at time $t_i$. Then
\begin{eqnarray}
\label{eqn13}
h(t;t_0) &  \hspace{-2mm} = &  \hspace{-2mm} (h_1 + h_2 e^{j 2 \pi \nu_2 t_0}) \delta(t - 2 \mu s)  \nonumber \\
 & & + \, h_3  \delta(t - 3 \mu s) \, + \, h_4 e^{j 2 \pi \nu_4 t_0}\delta(t - 4 \mu s),
\end{eqnarray}
and
\begin{eqnarray}
\label{eqn14}
h(t;t_1) & \hspace{-2mm} = &  \hspace{-2mm} (h_1 + h_2 e^{j 2 \pi \nu_2 t_1}) \delta(t - 2 \mu s)  \nonumber \\
 & & + \, h_3  \delta(t - 3 \mu s) \, + \, h_4 e^{j 2 \pi \nu_4 t_1}\delta(t - 4 \mu s).
\end{eqnarray}
We look to predict the received TD signal
$y_1(t) =  h(t - t_1 ; t_1)$
 from the previous signal $y_0(t) =  h(t - t_0 ; t_0)$.
 After estimating the path delays from $y_0(t)$, we can predict that $y_1(t)$ will involve path delays of
 $2 \, \mu s$, $3 \, \mu s$ and $4 \, \mu s$. 
 
 {$\boldsymbol{h_3 \delta(t - 3 \, \mu s)}$}: This term due to the third channel path is
 common to $h(t ; t_0)$ and $h(t ; t_1)$.
 \emph{Channel paths which do not induce Doppler shift and whose delay is distinct from that of other paths result in a predictable channel interaction.} The complex gain $h_3$ in this term does not depend on when the pulse was transmitted, so the interaction of the pulse with this
 path is stationary and non-fading (received pulse power is independent of when the pulse was transmitted). The response of such a channel path to any transmitted pulse can therefore be \emph{predicted}.  
 
 {$\boldsymbol{h_4 e^{j 2 \pi \nu_4 t_0} \delta(t - 4 \, \mu s)}$ and $\boldsymbol{h_4 e^{j 2 \pi \nu_4 t_1} \delta(t - 4 \, \mu s)}$}:
 These terms are due to the fourth channel path, whose Doppler shift $\nu_4$ interacts with the time at which the pulse was transmitted. As it is not possible to simultaneously estimate $h_4$ and $\nu_4$ from $h_4 e^{j 2 \pi \nu_4 t_0}$, we cannot predict the complex gain $h_4 e^{j 2 \pi \nu_4 t_1}$
 of the corresponding term in $h(t; t_1)$.
 The received signal power is time-independent since $\vert h_4 e^{j 2 \pi \nu_4 t_i} \vert^2 = \vert h_4 \vert^2$ but the phase is time-dependent.
 \emph{Channel paths, which induce Doppler shift and whose delay is different from that of other paths, result in a non-predictable channel interaction which is non-stationary and non-fading.}
 
${\boldsymbol{\left(h_1 + h_2 e^{j 2 \pi \nu_2 t_0} \right) \delta(t - 2 \, \mu s)}}$ and ${\boldsymbol{\left(h_1 + h_2 e^{j 2 \pi \nu_2 t_1} \right) \delta(t - 2 \, \mu s)}}$: The first and second paths have the same delay, but only the second path introduces a Doppler shift. It is not possible to predict
${\left(h_1 + h_2 e^{j 2 \pi \nu_2 t_1} \right) \delta(t - 2 \, \mu s)}$ since it is not possible to separately estimate $h_1, h_2, \nu_1$ and $\nu_2$ from ${\left(h_1 + h_2 e^{j 2 \pi \nu_2 t_0} \right) \delta(t - 2 \, \mu s)}$. Note that in this case both the complex gain and received signal power are time-dependent (i.e., they depend on the time at which the pulse is transmitted).
\emph{Two or more channel paths which have the same path delay and at least one of which induces a Doppler shift which is different from that induced by the other paths, result in
non-predictable channel interaction which is fading and non-stationary}.

\subsection{Interaction with a FD pulse}
\label{secChannelInteractionFD}
We consider the same stationary doubly-spread channel
illustrated in Fig.~\ref{fig_7}. Here we consider
the FD response to a transmitted FD signal given by
\begin{eqnarray}
\label{eqn14}
X(f) & = & \delta(f - f_0) \, + \, \delta(f - f_1)
\end{eqnarray}
comprising two FD pulses transmitted at $f = f_0 = 15$ KHz
and $f = f_1 = 450$ KHz.
We let $H(f; f_i)$ denote the FD impulse response of the channel to a FD impulse transmitted at frequency $f_i$. Then
\begin{eqnarray}
\label{eqn20}
H(f; f_0) & \hspace{-2mm} = & \hspace{-2mm} {\Big [} h_1 \, e^{- j 2 \pi f_0 \tau_1} \, + \, h_3 \, e^{- j 2 \pi f_0 \tau_3} {\Big ]} \delta(f) \nonumber \\
& & + \, h_2 \, e^{-j 2 \pi (f_0 + \nu_2) \tau_2 } \, \delta(f  + 950) \nonumber \\
& & + \, h_4 \, e^{-j 2 \pi (f_0 + \nu_4) \tau_4 } \, \delta(f  - 750)
\end{eqnarray}
and
\begin{eqnarray}
\label{eqn21}
H(f; f_1) & \hspace{-2mm} = & \hspace{-2mm} {\Big [} h_1 \, e^{- j 2 \pi f_1 \tau_1} \, + \, h_3 \, e^{- j 2 \pi f_1 \tau_3} {\Big ]} \delta(f) \nonumber \\
& & + \, h_2 \, e^{-j 2 \pi (f_1 + \nu_2) \tau_2 } \, \delta(f  + 950) \nonumber \\
& & + \, h_4 \, e^{-j 2 \pi (f_1 + \nu_4) \tau_4 } \, \delta(f  - 750).
\end{eqnarray}
Again we look to predict the received FD signal $Y_1(f) = H(f - f_1; f_1)$ due to a pulse transmitted at $f=f_1$ from the received signal $Y_0(f) = H(f - f_0; f_0)$ due to the pulse transmitted at $f=f_0$.
Since the underlying channel is stationary, we can accurately predict that $H(f; f_1)$ will comprise impulses at $0$ Hz, $-950$ Hz and $750$ Hz. 

{$\boldsymbol{h_2 \, e^{- j 2 \pi (f_0 + \nu_2) \tau_2} \, \delta(f + 950 \, Hz)}$}: The second path interacts with the FD pulse transmitted at $f=f_0$ resulting in a received pulse at $f= f_0 + \nu_2$ having complex gain $h_2 \, e^{- j 2 \pi (f_0 + \nu_2) \tau_2}$. From this complex gain we cannot predict the complex gain $h_2 \, e^{- j 2 \pi (f_1 + \nu_2) \tau_2}$ of the corresponding term in $Y_1(f)$
since it is not possible to estimate $h_2$ and $\tau_2$
from $h_2 \, e^{- j 2 \pi (f_0 + \nu_2) \tau_2}$. Since there is no other path with the same Doppler shift, the magnitude of the
received FD pulse is $\vert h_2 \, e^{-j 2 \pi (f_i + \nu_2) \tau_2 } \vert = \vert h_2 \vert$. There is no FD fading
since this magnitude does not depend on the FD location of the transmitted pulse. The same analysis applies to the interaction
of the $4$th channel path with the FD pulses. \emph{Channel paths whose Doppler shift is distinct from that of other paths result in a non-predictable, non-fading and non-stationary
interaction between the channel and the FD pulse}.

{$\boldsymbol{\left( h_1  e^{- j 2 \pi f_0 \tau_1 } \, + \, h_3 e^{- j 2 \pi f_0 \tau_3 } \right)  \delta(f)}$}: The $1$st and $3$rd paths do not induce any Doppler shift and have different path delays. The interaction of these two paths with the pulse transmitted at $f=f_0$ results in a received pulse at $f=f_0$ having complex gain $\left( h_1  e^{- j 2 \pi f_0 \tau_1 } \, + \, h_3 e^{- j 2 \pi f_0 \tau_3 } \right)$. From this complex gain, it is not possible to predict the complex gain $\left(  h_1  e^{- j 2 \pi f_1 \tau_1 } \, + \, h_3 e^{- j 2 \pi f_1 \tau_3 }\right)$ of the pulse received due to the interaction of the 1st and 3rd paths with the pulse transmitted at $f=f_1$.  
This is because, it is not possible to separately estimate $h_1, h_3, \tau_1$ and $\tau_3$ from $\left(  h_1  e^{- j 2 \pi f_0 \tau_1 } \, + \, h_3 e^{- j 2 \pi f_0 \tau_3 }\right)$.
This interaction is non-stationary and fading. \emph{Two or more channel paths which induce the same Doppler shift, and at least one of which has a path delay which is different from that of the other paths result in a non-predictable
channel interaction which is fading and non-stationary.}

\subsection{Interaction with a DD domain pulse}
\label{secChannelInteractionDD}
\begin{figure*}[!h]
\centering
\includegraphics[width=16cm, height=8.0cm]{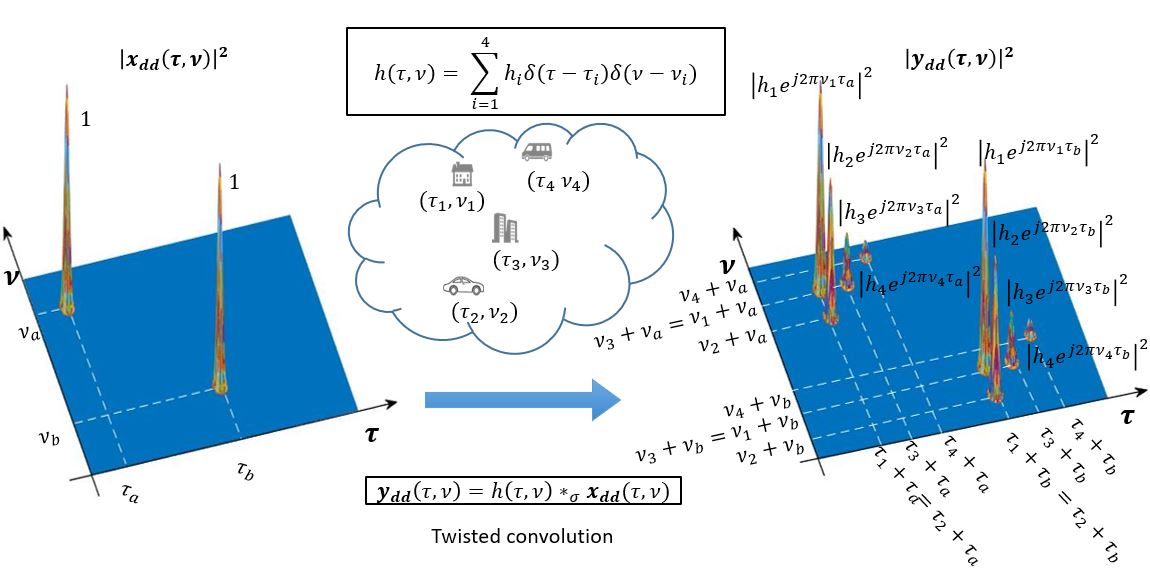}
\caption{The transmit DD pulse at $(\tau_a, \nu_a)$ is ``effectively" localized along both delay and Doppler domain, and therefore DD pulses are received along each path at distinct DD locations $(\tau_a + \tau_i, \nu_a + \nu_i), i=1,2,3,4$ which allows for accurate estimation of the channel DD spreading function $h(\tau,\nu)$ (see (\ref{eqn1})). Due to quasi-periodicity, the transmit DD pulse is repeated along the delay and Doppler domain with period $\tau_p$ and $\nu_p$, respectively. These repetitions also interact with the channel resulting in received DD pulses. If $\tau_p$ is less than the channel delay spread and/or if $\nu_p$ is less than the channel Doppler spread (i.e, $(\tau_p, \nu_p)$ does not satisfy (\ref{eqn28})), then the received DD pulses corresponding to the quasi-periodic pulse repetition could overlap/alias with the pulses received due to the DD pulse transmitted at $(\tau_a, \nu_a) \in {\mathcal D}_0$, thereby making channel prediction non-simple and difficult. 
However, if $(\tau_p, \nu_p)$ is properly chosen to satisfy (\ref{eqn28}) then the channel response
to a DD pulse transmitted at some other DD location (e.g., $(\tau_b, \nu_b)$) can be predicted from perfect knowledge of the channel response to the DD pulse transmitted at $(\tau_a, \nu_a)$. Also, when $(\tau_p,\nu_p)$ satisfies (\ref{eqn28}), the energy of the received DD domain pulses is invariant of the location of the transmitted DD pulse. Therefore, the interaction of a DD pulse with the channel is \emph{predictable and non-fading} if (\ref{eqn28}) is satisfied.}
\label{fig_8}
\end{figure*}
We continue to study the stationary doubly-spread channel illustrated in Fig.~\ref{fig_7}.
In this section, we analyze the interaction of this channel with two DD domain pulses
transmitted at $(\tau,\nu) = (\tau_a, \nu_a)$ and $(\tau,\nu) = (\tau_b, \nu_b)$, $0 \leq \tau_a, \tau_b < \tau_p$, $0 \leq \nu_a, \nu_b < \nu_p$. From (\ref{eqn4}) it follows that the
corresponding quasi-periodic DD domain signal is given by

{\vspace{-4mm}
\small
\begin{eqnarray}
\label{eqn22}
x_{_{\mbox{\footnotesize{dd}}}}(\tau,\nu) & \hspace{-2mm} = & \hspace{-2mm} \sum\limits_{m \in {\mathbb Z}}\sum\limits_{n \in {\mathbb Z}} e^{j 2 \pi n \nu \tau_p } {\Big [}  \delta(\tau - n \tau_p - \tau_a) \delta (\nu - m \nu_p - \nu_a) \nonumber \\
& & \,\,\,\,\,\,\,\, +  \, \delta(\tau - n \tau_p - \tau_b) \delta (\nu - m \nu_p - \nu_b) {\Big ]}.
\end{eqnarray}
\normalsize}
We obtain the transmitted TD signal $x(t)$ by applying the inverse Zak transform
${\mathcal Z}_t^{-1}$ as in (\ref{eqn6}).
\begin{eqnarray}
\label{eqn23}
x(t) & = & \sqrt{\tau_p} \int_{0}^{\nu_p} x_{_{\mbox{\footnotesize{dd}}}}(t, \nu) \, d\nu \nonumber \\
& = & \sqrt{\tau_p} \sum\limits_{n \in {\mathbb Z}} e^{j 2 \pi n \nu_a \tau_p}
\, \delta(t - \tau_a - n \tau_p) \nonumber \\
& & \, + \, \sqrt{\tau_p} \sum\limits_{n \in {\mathbb Z}} e^{j 2 \pi n \nu_b \tau_p}
\, \delta(t - \tau_b - n \tau_p).
\end{eqnarray}
We obtain the received TD signal $y(t)$ by substituting (\ref{eqn23}) in (\ref{eqn3}).

{\vspace{-4mm}
\small
\begin{eqnarray}
\label{eqn24}
y(t) & \hspace{-3mm} = & \hspace{-3mm} \sqrt{\tau_p} \sum\limits_{i=1}^{4}  h_i {\Big [} e^{j 2 \pi \nu_i \tau_a}  \sum\limits_{n \in {\mathbb Z}} e^{j2 \pi n (\nu_a + \nu_i) \tau_p}  \delta(t - (\tau_a + \tau_i) - n \tau_p)  \nonumber \\
& & \,\,\,\, + \, e^{j 2 \pi \nu_i \tau_b} \sum\limits_{n \in {\mathbb Z}} e^{j2 \pi n (\nu_b + \nu_i) \tau_p}  \delta(t - (\tau_b + \tau_i) - n \tau_p) {\Big ]} .
\end{eqnarray}
\normalsize}
We obtain $y_{_{\mbox{\footnotesize{dd}}}}(\tau,\nu)$, the DD domain representation of
$y(t)$ by applying the Zak transform ${\mathcal Z}_t$ as in (\ref{eqn7}). We write
\begin{eqnarray}
\label{eqn25}
y_{_{\mbox{\footnotesize{dd}}}}(\tau,\nu) & = & y_{_{\mbox{\footnotesize{dd}}}}(\tau,\nu ; \tau_a, \nu_a) \, + \, y_{_{\mbox{\footnotesize{dd}}}}(\tau,\nu ; \tau_b, \nu_b),
\end{eqnarray}
where

{\vspace{-4mm}
\small
\begin{eqnarray}
\label{eqn26}
y_{_{\mbox{\footnotesize{dd}}}}(\tau,\nu ; \tau_a, \nu_a) & \hspace{-3mm} = & \hspace{-3mm} \sum\limits_{i=1}^4 h_i e^{j 2 \pi \nu_i \tau_a} {\Big [} \hspace{-2mm} \sum\limits_{m,n \in {\mathbb Z}} \hspace{-2mm} e^{j 2 \pi n \nu \tau_p}  \delta(\tau - \tau_a - \tau_i - n \tau_p) \nonumber \\
& & \hspace{19mm} \delta(\nu - \nu_a - \nu_i - m \nu_p) {\Big ]}
\end{eqnarray}\normalsize}
\hspace{-2.5mm}
{is the response associated with the pulse transmitted at $(\tau_a, \nu_a)$, referred to as the $a$-response, and}

{\vspace{-4mm}
\small
\begin{eqnarray}
\label{eqn27}
y_{_{\mbox{\footnotesize{dd}}}}(\tau,\nu ; \tau_b, \nu_b) & \hspace{-3mm} = & \hspace{-3mm} \sum\limits_{i=1}^4 h_i e^{j 2 \pi \nu_i \tau_b} {\Big [} \hspace{-2mm} \sum\limits_{m,n \in {\mathbb Z}} \hspace{-2mm} e^{j 2 \pi n \nu \tau_p}  \delta(\tau - \tau_b - \tau_i - n \tau_p) \nonumber \\
& & \hspace{19mm} \delta(\nu - \nu_b - \nu_i - m \nu_p) {\Big ]}
\end{eqnarray}\normalsize}
\hspace{-2.5mm}
{is the response associated with the pulse transmitted at $(\tau_b, \nu_b)$, referred to as the $b$-response.}
Recall from Section \ref{secDDCarrier}, that channel path delay shifts the DD domain pulse
along the delay axis, and that channel path Doppler shifts the DD domain pulse
along the Doppler axis.
{From (\ref{eqn26}), we see that
the $a$-response includes four distinct pulses received at $(\tau_a + \tau_i, \nu_a + \nu_i)$, and from (\ref{eqn27}), we see that the $b$-response includes four distinct pulses received at $(\tau_b + \tau_i, \nu_b + \nu_i)$, where $i=1,2,3,4$.} Fig.~\ref{fig_8} illustrates that the received pulse along each path
is observed \emph{separately} in the DD domain.

Again we look to predict the received DD domain signal $y_{_{\mbox{\footnotesize{dd}}}}(\tau,\nu ; \tau_b, \nu_b)$ from the signal $y_{_{\mbox{\footnotesize{dd}}}}(\tau,\nu ; \tau_a, \nu_a)$.\\

\underline{The channel interaction is predictable}: 
{\emph{The channel interaction is predictable when the delay period $\tau_p$ is greater than the channel delay spread and the Doppler period $\nu_p$ is greater than the channel Doppler spread,}} i.e.,
    \begin{eqnarray}
    \label{eqn28}
    \tau_p & > & (\max_i \tau_i  \, - \, \min_i \tau_i), \nonumber \\
    \mbox{\small{and}} \,\,\,\,  \nu_p & > & (\max_i \nu_i  \, - \, \min_i \nu_i).    \end{eqnarray}
{Condition (\ref{eqn28}) is called the \emph{crystallization condition}. In our example, when the crystallization condition holds, the $b$-response can be predicted from the $a$-response.
Fig.~\ref{fig_8} illustrates this fact. It shows that the complex gain of the pulse received along the $i$th path is $h_i e^{j 2 \pi \nu_i \tau_a}$ for the $a$-response, and $h_i e^{j 2 \pi \nu_i \tau_b}$ for the $b$-response. The later can be predicted from the former via}  
\begin{eqnarray}
\label{eqn28p5}
\hspace{-4mm}
h_i e^{j 2 \pi \nu_i \tau_b} & \hspace{-2mm} = & \hspace{-2mm} h_i e^{j 2 \pi \nu_i \tau_a} \,  e^{j 2 \pi \nu_i (\tau_b - \tau_a)}, \ \ i=1,2,3,4.
\end{eqnarray}

\begin{figure*}[!h]
\centering
\includegraphics[width=16cm, height=8.0cm]{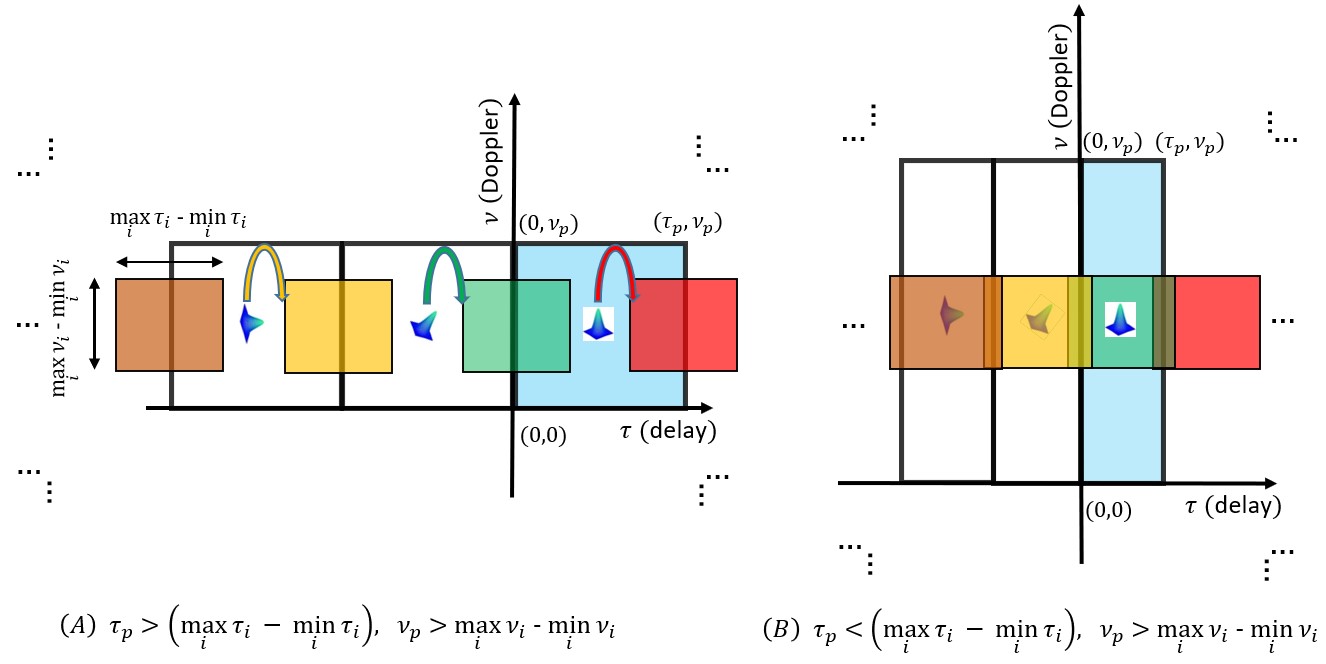}
\caption{
A single DD domain pulse is transmitted in the fundamental DD domain period ${\mathcal D}_0$.
Rectangular DD domain regions containing the received DD domain pulses corresponding to the transmitted DD pulse in ${\mathcal D}_0$ and its quasi-periodic repetitions are depicted as colour filled rectangles.
The size of each such rectangular region is invariant of the delay and Doppler domain period, i.e., they have length equal to the channel delay spread $\left( \max_i \tau_i \, - \, \min_i \tau_i \right)$ along the delay axis and equal to the channel Doppler spread $\left( \max_i \nu_i \, - \, \min_i \nu_i \right)$ along the Doppler axis. Two choices of the delay and Doppler domain period are considered, choice A: the delay and Doppler domain period satisfy the condition in (\ref{eqn28}), and
choice B: Doppler domain period is greater than the channel Doppler spread but the delay domain period is \emph{less} than the channel delay spread.
The fundamental DD domain period ${\mathcal D}_0$ is depicted by a blue filled rectangle. For choice A, DD pulses received in ${\mathcal D}_0$ are within the green and the red coloured rectangles which do not overlap and therefore there is no aliasing.
However, in choice B, due to small $\tau_p$ and the invariance of the size of the coloured rectangles, the green rectangle overlaps with the red and the yellow coloured rectangles resulting in delay domain aliasing.  
Similarly, there would be aliasing along the Doppler domain if the Doppler domain period is chosen to be smaller than the channel Doppler spread.} 
\label{fig_ddalias}
\end{figure*}

{The predictive relation (\ref{eqn28p5}) breaks down when responses
associated with replicas outside the fundamental period interacts with responses associated with replicas inside the fundamental period. This phenomenon is referred to as DD domain aliasing.
The point is that when the crystallization condition holds, DD domain aliasing is precluded. Here is an example of an aliasing situation: say $\tau_p = \tau_4 - \tau_1$ and $\nu_p = \nu_4 - \nu_1$. In this situation, the responses associated with the pulse transmitted at $(\tau_a,\nu_a)$ and its replica at $(\tau_a-\tau_p, \nu_a - \nu_p)$ are both received at the same location $(\tau_a + \tau_1, \nu_a + \nu_1)$
along the $1$st and the $4$th channel paths, respectively.
At this location, the complex gain of the $a$-response is
$\left( h_1 e^{j 2 \pi \nu_1 \tau_a} + h_4 \, e^{j 2 \pi \nu_4 \tau_a} \, e^{-j 2 \pi (\nu_a + \nu_1)(\tau_4 - \tau_1)}\right)$,
from which it is difficult to separately estimate the terms corresponding to $h_1$ and $h_4$.
Therefore, the complex gain $\left( h_1 e^{j 2 \pi \nu_1 \tau_b} + h_4 e^{j 2 \pi \nu_4 \tau_b} e^{-j 2 \pi (\nu_b + \nu_1)(\tau_4 - \tau_1)}\right)$ of the received pulse at $(\tau_b + \tau_1, \nu_b + \nu_1)$ in
the $b$-response cannot be simply predicted from the complex gain of the received pulse at $(\tau_a + \tau_1, \nu_a + \nu_1)$ in the $a$-response. Fig.~\ref{fig_ddalias} depicts the phenomena of DD domain aliasing.}   

Conditions (\ref{eqn28}) preclude aliasing from the replicas of the DD domain pulse
transmitted in ${\mathcal D}_0$.
Since $\tau_p \, \nu_p = 1$, we require
\begin{eqnarray}
\label{eqn31}
(\max_i \tau_i  \, - \, \min_i \tau_i) \, \times \, (\max_i \nu_i  \, - \, \min_i \nu_i) & < & 1,
\end{eqnarray}
that is the product of the channel delay spread and Doppler spread is less than one.
This condition is generally satisfied for most doubly-spread channels of practical interest. For example, in a typical cellular wireless system with channel delay spread $5 \mu s$ and Doppler spread $1000$ Hz, the product of the delay and Doppler spread is only $5 \times 10^{-3}$.\\

\underline{The channel interaction is non-fading}:
{\emph{The channel interaction is non-fading when the crystallization condition (\ref{eqn28}) holds, in the sense that the amplitude of the response does not depend on the location of the transmitted pulse.} Specifically, in our example, when the crystallization condition holds, the power distribution of the $b$-response is equal to the power distribution of the $a$-response. Fig.~\ref{fig_8} illustrates this fact.
It shows that in both distributions the amplitude of the pulse received along the $i$-th path is $\vert h_i \vert$. When the
crystallization condition does not hold, due to aliasing effect,
the power distribution in general depends on the location of the
transmit pulse. We demonstrate this using the aliasing example considered in the predictability discussion. In this example, the
amplitude of the pulse received at $(\tau_a + \tau_1 , \nu_a + \nu_1)$ is $\vert \left( h_1 e^{j 2 \pi \nu_1 \tau_a} + h_4 \, e^{j 2 \pi \nu_4 \tau_a} \, e^{-j 2 \pi (\nu_a + \nu_1)(\tau_4 - \tau_1)}\right) \vert$. It can be verified that this expression
in general depends on the location $(\tau_a, \nu_a)$ of the transmitted pulse.}

\underline{The channel interaction is non-stationary}:
{It should be noted that even when the crystallization condition holds, the actual phases in the response depend on the location of the transmitted pulse. In our example,
the complex gain of the pulse received along the $i$th path is
$h_i e^{j 2 \pi \nu_i \tau_a}$ in the $a$-response and is $h_i e^{j 2 \pi \nu_i \tau_b}$ in the $b$-response. The point is that
non-stationarity is not a major issue as long as predictability
(\ref{eqn28p5}) is maintained.}
{
\begin{figure*}[!h]
\centering
\vspace{-7mm}
\includegraphics[width=19cm, height=5.4cm]{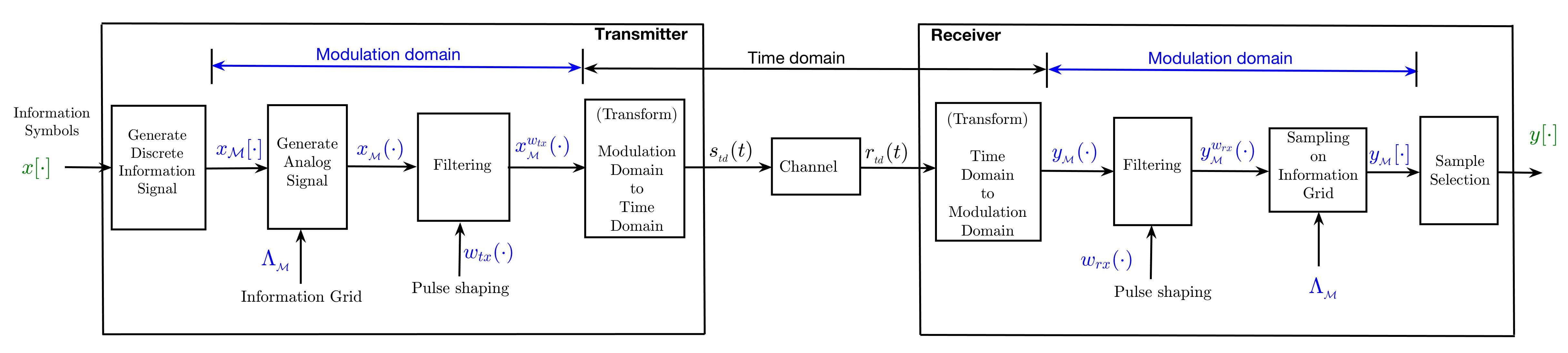}
\caption{Transceiver signal processing.}
\label{fig_8b}
\end{figure*}}

\section{Transceiver Signal Procesing}
\label{iorelation}
In this section, we show that OTFS
is similar to TDM and FDM in that it
involves a similar sequence of signal processing operations at the transmitter and receiver.
In TDM and FDM respectively, information is embedded in TD and the FD, whereas
in OTFS it is embedded in the DD domain.

Throughout this section, we consider transmitting a packet of $BT$ information symbols $x[k], k=0,1,\cdots, (BT-1)$ using
a TD signal limited to time $T$ and bandwidth $B$.
The information symbols are first converted to a discrete information signal $x_{_{\mathcal M}}[\cdot]$ in the \emph{modulation domain} ${\mathcal M}$.
The subscripts \emph{td}, \emph{fd} and \emph{dd} identify the TD, FD and DD modulation domains. 
Within the modulation domain, we then
apply a filter $w_{tx}(\cdot)$ to ensure that the transmit TD information signal $s_{\mbox{\footnotesize{td}}}(t)$ satisfies the time and bandwidth constraints.
The signal $s_{\mbox{\footnotesize{td}}}(t)$ interacts with the doubly-spread channel to provide a received TD signal $r_{\mbox{\footnotesize{td}}}(t)$.
This signal is then converted to the appropriate modulation domain, using the inverse of the transform used at the transmitter, to provide a signal $y_{_{\mathcal M}}(\cdot)$.
Matched filtering at the receiver with the receive filter $w_{rx}(\cdot)$ then optimizes the
signal to noise ratio (SNR) in the post-filtered signal to produce a modulation domain analog signal $y_{_{\mathcal M}}^{w_{rx}}(\cdot)$.
This signal is then sampled to provide the discrete received modulation domain signal $y_{_{\mathcal M}}[\cdot]$ which is processed to recover the information symbols.

These steps are shown in Fig.~\ref{fig_8b}
and are made explicit in Table-\ref{tabTDMFDM} for TDM and FDM. Table-\ref{tab1} lists the transforms, filtering operation mentioned in Fig.~\ref{fig_8b} for TDM/FDM/OTFS. 
\begin{table*}
\caption{Signal processing steps in TDM and FDM.}
\centering
\begin{tabular}{ | c || c |  c | } 
  \hline
   Transceiver operation & TDM & FDM   \\
   \hline
  Generating the discrete & 
  $\begin{aligned}
x_{\mbox{\footnotesize{td}}}[k] & \Define \begin{cases}
x[k] &, k=0,1,\cdots,(BT-1) \\
0 &, \mbox{\small{otherwise}}
\end{cases}. \nonumber
\end{aligned}$ &  $\begin{aligned}
x_{\mbox{\footnotesize{fd}}}[k] & \Define \begin{cases}
x[k] &, k=0,1,\cdots,(BT-1) \\
0 &, \mbox{\small{otherwise}}
\end{cases}. \nonumber
\end{aligned}$  \\ 
  information signal & &   \\
  \hline
Generating the analog signal & ${\Lambda}_{\mbox{\footnotesize{td}}} = \{ q/B \, | \, q \in {\mathbb Z}\}$ &   ${\Lambda}_{\mbox{\footnotesize{fd}}} = \{ q/T \, | \, q \in {\mathbb Z}\}$ \\
on the information grid & $\begin{aligned}
x_{\mbox{\footnotesize{td}}}(t) & = & \sum\limits_{k \in {\mathbb Z}} x_{\mbox{\footnotesize{td}}}[k] \, \delta(t - k/B)
\end{aligned}$
&  $\begin{aligned}
x_{\mbox{\footnotesize{fd}}}(f)  =  \sum\limits_{k \in {\mathbb Z}} x_{\mbox{\footnotesize{fd}}}[k] \, \delta(f - k/T)
\end{aligned}$ \\
\hline
Shaping the pulse & 
$\begin{aligned}
x_{\mbox{\footnotesize{td}}}^{w_{tx}}(t)  = w_{tx}(t) \, * \, x_{\mbox{\footnotesize{td}}}(t) \nonumber
\end{aligned}$
& $\begin{aligned}
x_{\mbox{\footnotesize{fd}}}^{w_{tx}}(f)  = w_{tx}(f) \, * \, x_{\mbox{\footnotesize{fd}}}(f) \nonumber
\end{aligned}$ \\
at the transmitter & & \\
\hline
Converting from the modulation & $s_{\mbox{\footnotesize{td}}}(t) = x_{\mbox{\footnotesize{td}}}^{w_{tx}}(t)$ & $s_{\mbox{\footnotesize{td}}}(t) = \int x_{\mbox{\footnotesize{fd}}}^{w_{tx}}(f) \, e^{j 2 \pi f t} \, df$\\
domain to the time domain & & \\
\hline
Applying the channel $h(\tau,\nu)$ &  $ r_{\mbox{\footnotesize{td}}}(t) = \iint h(\tau,\nu) s_{\mbox{\footnotesize{td}}}(t-\tau) \, e^{j 2 \pi \nu (t - \tau)} \, d\tau \, d\nu$ & $ r_{\mbox{\footnotesize{td}}}(t) = \iint h(\tau,\nu) s_{\mbox{\footnotesize{td}}}(t-\tau) \, e^{j 2 \pi \nu (t - \tau)} \, d\tau \, d\nu$ \\
\hline
Converting from the time domain & $y_{\mbox{\footnotesize{td}}}(t) = r_{\mbox{\footnotesize{td}}}(t)$ &  $y_{\mbox{\footnotesize{fd}}}(f) = \int r_{\mbox{\footnotesize{td}}}(t) \, e^{-j 2 \pi f t} \, dt$\\
to modulation domain &  &  \\
\hline
Shaping the pulse & $y_{\mbox{\footnotesize{td}}}^{w_{rx}}(t) = w_{rx}(t) * y_{\mbox{\footnotesize{td}}}(t)$ & {$y_{\mbox{\footnotesize{fd}}}^{w_{rx}}(f) = w_{rx}(f) * y_{\mbox{\footnotesize{fd}}}(f)$} \\
at the receiver &  & \\
\hline
Sampling on the &  $\begin{aligned}
y_{\mbox{\footnotesize{td}}}[k]  =  y_{\mbox{\footnotesize{td}}}^{w_{rx}}(t = {k}/{B})\,,\, k \in {\mathbb Z}.
\end{aligned}$
&  $\begin{aligned}
y_{\mbox{\footnotesize{fd}}}[k]  =  y_{\mbox{\footnotesize{fd}}}^{w_{rx}}(f = {k}/{T} )\,,\, k \in {\mathbb Z}.
\end{aligned}$\\
information grid &  & \\
\hline
\end{tabular}
\label{tabTDMFDM}
\end{table*}

\begin{table}
\caption{Signal processing operations for TDM/FDM/OTFS in Fig.~\ref{fig_8b}.}
\centering
\begin{tabular}{ | c || c |  c |  c | } 
  \hline
   & TDM & FDM & OTFS  \\
   \hline
  Modulation & TD & FD & DD  \\ 
  domain & & &  \\
  \hline
  Transform & Identity & Fourier & Zak \\
  \hline
  Filtering & Linear & Linear & Twisted \\
            & convolution & convolution & convolution \\
            & in TD & in FD & in DD \\
  \hline
\end{tabular}
\label{tab1}
\end{table}

\subsection{TDM input-output relation}
\label{secTDMFDMIOrelation}
It follows from Table-\ref{tabTDMFDM} that the output signal
$y_{\mbox{\footnotesize{td}}}[k], k \in {\mathbb Z}$ is obtained from the input signal $x_{\mbox{\footnotesize{td}}}[k], k \in {\mathbb Z}$ by a discrete time convolution

{\vspace{-4mm}
\small
\begin{eqnarray}
\label{eqn38}
y_{\mbox{\footnotesize{td}}}[k'] & = & \sum\limits_{k \in {\mathbb Z}} x_{\mbox{\footnotesize{td}}}[k] \, h_{_{\mbox{\footnotesize{td}}}}[k' - k \, ; \, k], \nonumber \\
& & \hspace{-25mm} \mbox{\small{where, for}} \, n \in {\mathbb Z}, \nonumber \\
h_{_{\mbox{\footnotesize{td}}}}[n \, ; \, k] &  \hspace{-3mm} \Define & \hspace{-3mm} \iiint {\Big [} e^{j 2 \pi \nu \frac{k}{B}} h(\tau,\nu) w_{rx}(\tau')  e^{j 2 \pi \nu (\frac{n}{B} - \tau - \tau')} \nonumber \\
& & \hspace{6mm} w_{tx}\left(\frac{n}{B} - \tau - \tau'\right) {\Big ]} \, d\tau \, d\nu \, d\tau'.
\end{eqnarray}
\normalsize}
The discrete TD filter $h_{_{\mbox{\footnotesize{td}}}}[ \cdot \, ; \, k]$ represents the effective discrete TD channel response
to the $k$th discrete input symbol $x_{\mbox{\footnotesize{td}}}[k]$.
The dependence of $h_{_{\mbox{\footnotesize{td}}}}[ \cdot \, ; \, k]$ on $k$ is not simple,  and knowledge of this filter
$h_{_{\mbox{\footnotesize{td}}}}[ \cdot \, ; \, q]$ for some integer $q$ is not sufficient to accurately predict $h_{_{\mbox{\footnotesize{td}}}}[ \cdot \, ; \, k]$
for all $k \in {\mathbb Z}$. Also, the expected received signal power ${\mathbb E}\left[ \left \vert  y_{\mbox{\footnotesize{td}}}[k'] \right \vert^2\right]$ varies with $k'$. Therefore, for a generic doubly-spread channel, the TDM input-output relation is \emph{non-predictable},
\emph{fading} and \emph{non-stationary} (cf. Section \ref{secChannelInteractionTD}). 

If there is no Doppler shift, then $h(\tau, \nu) = g(\tau) \delta(\nu)$. Let $h_{_{\mbox{\footnotesize{td}}}}(t) \Define w_{rx}(t) * g(t) * w_{tx}(t)$, then substituting $h(\tau, \nu) = g(\tau) \delta(\nu)$ in (\ref{eqn38}) we get
$h_{_{\mbox{\footnotesize{td}}}}[ n \, ; \, k]= h_{_{\mbox{\footnotesize{td}}}}(t = n/B)$ which is independent of $k$. In this case, the TDM input-output relation is predictable, non-fading and stationary.
These facts are summarized in Table-\ref{tab2}.

\subsection{FDM input-output relation}
\label{secFDMIOrelation}
We see from Table-\ref{tabTDMFDM} that
the output signal $y_{\mbox{\footnotesize{fd}}}[k], k \in {\mathbb Z}$
is obtained from the input signal $x_{\mbox{\footnotesize{fd}}}[k], k \in {\mathbb Z}$ by the discrete FD convolution 

{\vspace{-4mm}
\small
\begin{eqnarray}
\label{eqn45}
y_{\mbox{\footnotesize{fd}}}[k'] & = & \sum\limits_{k \in {\mathbb Z}} x_{\mbox{\footnotesize{fd}}}[k] \, h_{_{\mbox{\footnotesize{fd}}}}[ k' - k \, ; \, k]  \nonumber \\
& & \hspace{-25mm} \mbox{\small{where, for}} \, n \in {\mathbb Z}, \nonumber \\
h_{_{\mbox{\footnotesize{fd}}}}[ n \, ; \, k] &  \hspace{-3mm} = & \hspace{-3mm} \iiint {\Big [} e^{-j 2 \pi \tau \frac{k}{T}} h(\tau,\nu) w_{rx}(f')  e^{-j 2 \pi \tau (\frac{n}{T} - f')} \nonumber \\
& & \hspace{5mm} w_{tx}\left(\frac{n}{T} - \nu - f'\right) {\Big ]} \, d\tau \, d\nu \, df'.
\end{eqnarray}
\normalsize}
The discrete FD filter $h_{_{\mbox{\footnotesize{fd}}}}[ \cdot \, ; \, k]$ represents the effective discrete FD channel response
to the $k$th discrete input symbol $x_{\mbox{\footnotesize{fd}}}[k]$.
The dependence of $h_{_{\mbox{\footnotesize{fd}}}}[ \cdot \, ; \, k]$ on $k$ is not simple,  and knowledge of this filter
$h_{_{\mbox{\footnotesize{fd}}}}[ \cdot \, ; \, q]$ for some integer $q$ is not sufficient to accurately predict $h_{_{\mbox{\footnotesize{fd}}}}[ \cdot \, ; \, k]$
for all $k \in {\mathbb Z}$. Also, the expected received signal power ${\mathbb E}\left[ \left \vert  y_{\mbox{\footnotesize{fd}}}[k'] \right \vert^2\right]$ varies with $k'$. Therefore, for a generic doubly-spread channel, the FDM input-output relation is \emph{non-predictable},
\emph{fading} and \emph{non-stationary} (cf. Section \ref{secChannelInteractionFD}).

If there are no path delays (Doppler only channel), then $h(\tau, \nu) = \delta(\tau) g(\nu)$. Let $h_{_{\mbox{\footnotesize{fd}}}}(f) \Define w_{rx}(f) * g(f) * w_{tx}(f)$, then substituting $h(\tau, \nu) = \delta(\tau) g(\nu)$ in (\ref{eqn45}) we get
$h_{_{\mbox{\footnotesize{fd}}}}[ n \, ; \, k]= h_{_{\mbox{\footnotesize{fd}}}}(f = n/T)$ which is independent of $k$. In this special case, the FDM input-output relation is predictable, non-fading and stationary. These facts regarding FDM are summarized in Table-\ref{tab2}.
\begin{table*}
\caption{Signal processing steps in OTFS.}
\centering
\begin{tabular}{ | c || c | } 
  \hline
   Transceiver operation & OTFS   \\
   \hline
  Generating the discrete information signal & 
   $\begin{aligned}
x_{_{\mbox{\footnotesize{dd}}}}[k+nM,l+mN] &  \Define \begin{cases}
x[k,l] & \hspace{-3mm} , m = n=0 \\
x[k,l] \, e^{j 2 \pi n \frac{l}{N}} & \hspace{-3mm} , \mbox{\small{otherwise}}
\end{cases}.
\end{aligned}$  \\ 
  \hline
Generating the analog signal on the information grid  &  
${\Lambda}_{_{\mbox{\footnotesize{dd}}}} \, = \,  \left\{ \left( k \frac{\tau_p}{M} , l \frac{\nu_p}{N}  \right) \, {\Big |} \, k, l \in {\mathbb Z} \right\}$\\
 &  $\begin{aligned}
 x_{_{\mbox{\footnotesize{dd}}}}(\tau, \nu)  =  \sum\limits_{k,l \in {\mathbb Z}}  x_{_{\mbox{\footnotesize{dd}}}}[k,l] \, \delta\left(\tau  - k \frac{\tau_p}{M} \right) \delta\left(\nu - l \frac{\nu_p}{N} \right)
\end{aligned}$ \\
\hline
Shaping the pulse at the transmitter  & $x_{_{\mbox{\footnotesize{dd}}}}^{w_{tx}}(\tau, \nu) \, = \,  w_{tx}(\tau,\nu) \, *_{\sigma} \, x_{_{\mbox{\footnotesize{dd}}}}(\tau, \nu)$ \\
\hline
Converting from the modulation domain to the time domain  & $s_{_{\mbox{\footnotesize{td}}}}(t) \, = \,  {\mathcal Z}_t^{-1}\left( x_{_{\mbox{\footnotesize{dd}}}}^{w_{tx}}(\tau, \nu) \right)$ \\
\hline
Applying the channel $h(\tau,\nu)$ &  $ r_{\mbox{\footnotesize{td}}}(t) = \iint h(\tau,\nu) s_{\mbox{\footnotesize{td}}}(t-\tau) \, e^{j 2 \pi \nu (t - \tau)} \, d\tau \, d\nu$ \\
\hline
Converting from the time domain to modulation domain  & $y_{_{\mbox{\footnotesize{dd}}}}(\tau, \nu)  = {\mathcal Z}_t\left( r_{_{\mbox{\footnotesize{td}}}}(t)\right)$ \\
\hline
Shaping the pulse at the receiver  &  $y_{_{\mbox{\footnotesize{dd}}}}^{w_{rx}}(\tau, \nu) = w_{rx}(\tau,\nu) \, *_{\sigma} \,  y_{_{\mbox{\footnotesize{dd}}}}(\tau, \nu)$ \\
\hline
Sampling on the information grid  & $y_{_{\mbox{\footnotesize{dd}}}}[k',l'] = y_{_{\mbox{\footnotesize{dd}}}}^{w_{rx}}\left(\tau = k' \frac{\tau_p}{M}, \nu = l' \frac{\nu_p}{N} \right), \,\, k', l' \in {\mathbb Z}$ \\
\hline
\end{tabular}
\label{tabOTFS}
\end{table*}

\subsection{OTFS input-output relation}
\label{secOTFSIOrelation}
OTFS modulation is parameterized by integers $M \approx B \tau_p$ and $N \approx T \nu_p$. Since OTFS is a 2-D modulation, the information symbols are usually
arranged as a 2-D finite array $x[k,l], k=0,1,\cdots,M-1, l=0,1,\cdots, N-1$.
The discrete DD domain information signal $x_{_{\mbox{\footnotesize{dd}}}}[k',l'], k',l' \in {\mathbb Z}$ is then
defined as follows. For all $k=0,1,\cdots, M-1$ and all $l=0,1,\cdots, N-1$ we have
\begin{eqnarray}
\label{eqn46}
x_{_{\mbox{\footnotesize{dd}}}}[k+nM,l+mN] & \hspace{-2mm} \Define \begin{cases}
x[k,l] & \hspace{-3mm} , m = n=0 \\
x[k,l] \, e^{j 2 \pi n \frac{l}{N}} & \hspace{-3mm} , \mbox{\small{otherwise}}
\end{cases}
\end{eqnarray}
where $m,n \in {\mathbb Z}$.
The DD domain information grid is given by
\begin{eqnarray}
\label{eqn465}
{\Lambda}_{_{\mbox{\footnotesize{dd}}}} & \Define & \left\{ \left( k \frac{\tau_p}{M} , l \frac{\nu_p}{N}  \right) \, {\Big |} \, k, l \in {\mathbb Z} \right\}.
\end{eqnarray}
The discrete DD domain information signal is then lifted to the information
grid ${\Lambda}_{_{\mbox{\footnotesize{dd}}}}$, resulting in the continuous DD domain
analog information signal $x_{_{\mbox{\footnotesize{dd}}}}(\tau, \nu)$ which is given by
\begin{eqnarray}
\label{eqn47}
x_{_{\mbox{\footnotesize{dd}}}}(\tau, \nu) & \hspace{-3mm} \Define & \hspace{-3mm} \sum\limits_{k,l \in {\mathbb Z}}  x_{_{\mbox{\footnotesize{dd}}}}[k,l] \, \delta\left(\tau  - k \frac{\tau_p}{M} \right) \delta\left(\nu - l \frac{\nu_p}{N} \right).
\end{eqnarray}
Since $x_{_{\mbox{\footnotesize{dd}}}}[\cdot,\cdot]$ satisfies (\ref{eqn46}), it follows that $x_{_{\mbox{\footnotesize{dd}}}}(\tau, \nu)$ is quasi-periodic.
In order to satisfy the time and bandwidth constraints for the transmit TD signal, $x_{_{\mbox{\footnotesize{dd}}}}(\tau, \nu)$ is filtered (twisted convolution) in the DD domain with a DD domain filter $w_{tx}(\tau,\nu)$ resulting in the
quasi-periodic DD domain signal\footnote{\footnotesize{Twisted convolution between two DD domain functions $a(\tau,\nu)$ and $b(\tau,\nu)$ is given by $a(\tau,\nu) *_{\sigma} b(\tau,\nu) \Define \iint a(\tau',\nu') \, b(\tau - \tau', \nu - \nu') \, e^{j 2 \pi \nu'(\tau - \tau')} \, d\tau'd\nu'$. Unlike linear convolution, twisted convolution is
non-commutative, i.e., $a(\tau,\nu) *_{\sigma} b(\tau,\nu) \ne b(\tau,\nu) *_{\sigma} a(\tau,\nu)$. It is however associative, i.e., $a(\tau,\nu) *_{\sigma} [ b(\tau,\nu) *_{\sigma} c(\tau,\nu) ] = [ a(\tau,\nu) *_{\sigma}  b(\tau,\nu) ]  *_{\sigma} c(\tau,\nu) $.}}
\begin{eqnarray}
\label{eqn48}
x_{_{\mbox{\footnotesize{dd}}}}^{w_{tx}}(\tau, \nu) & \Define & w_{tx}(\tau,\nu) \, *_{\sigma} \, x_{_{\mbox{\footnotesize{dd}}}}(\tau, \nu).  
\end{eqnarray}
The inverse Zak transform (see (\ref{eqn6})) of this filtered signal gives the transmit TD signal $s_{_{\mbox{\footnotesize{td}}}}(t)$, i.e.,
\begin{eqnarray}
\label{eqn49}
s_{_{\mbox{\footnotesize{td}}}}(t) & \Define & {\mathcal Z}_t^{-1}\left( x_{_{\mbox{\footnotesize{dd}}}}^{w_{tx}}(\tau, \nu) \right).
\end{eqnarray}
The received TD signal $r_{_{\mbox{\footnotesize{td}}}}(t)$ is then given by
\begin{eqnarray}
\label{eqn35}
r_{_{\mbox{\footnotesize{td}}}}(t) & \hspace{-2mm} = &  \hspace{-2mm} \iint h(\tau,\nu) s_{\mbox{\footnotesize{td}}}(t-\tau) \, e^{j 2 \pi \nu (t - \tau)} \, d\tau \, d\nu.
\end{eqnarray}
At the receiver, Zak transform (\ref{eqn7}) converts the received TD signal to a DD domain signal $y_{_{\mbox{\footnotesize{dd}}}}(\tau, \nu)$, i.e.,
\begin{eqnarray}
\label{eqn50}
y_{_{\mbox{\footnotesize{dd}}}}(\tau, \nu) & \Define & {\mathcal Z}_t\left( r_{_{\mbox{\footnotesize{td}}}}(t)\right).
\end{eqnarray}
This received DD domain signal is then match-filtered with a DD domain receive filter $w_{rx}(\tau,\nu)$
resulting in the filtered DD domain signal
\begin{eqnarray}
\label{eqn51}
y_{_{\mbox{\footnotesize{dd}}}}^{w_{rx}}(\tau, \nu) & \Define & w_{rx}(\tau,\nu) \, *_{\sigma} \,  y_{_{\mbox{\footnotesize{dd}}}}(\tau, \nu).
\end{eqnarray}
This quasi-periodic signal is then sampled on the information grid, resulting in the discrete DD
domain received signal
\begin{eqnarray}
\label{eqn52}
y_{_{\mbox{\footnotesize{dd}}}}[k',l'] & \hspace{-3mm} \Define &  \hspace{-3mm} y_{_{\mbox{\footnotesize{dd}}}}^{w_{rx}}\left(\tau = k' \frac{\tau_p}{M}, \nu = l' \frac{\nu_p}{N} \right), \,\, k', l' \in {\mathbb Z}.
\end{eqnarray}
The signal processing steps from (\ref{eqn46})-(\ref{eqn52}) are summarized in Table-\ref{tabOTFS}.\footnote{\footnotesize{{The receiver processing, i.e., Zak transform of the received TD signal followed by twisted convolution with the receive DD filter and subsequent DD domain sampling can be implemented efficiently in the discrete DD domain using the Discrete Zak Transform (DZT  \cite{Lampel2022}) on the sampled received TD signal (sampled at integer multiples of $1/B$). For a given $(M,N)$, the complexity of DZT is $O(MN \log N)$ when compared to the $O(MN \log (MN))$ complexity of DFT/IDFT processing in a FDM based system with the same frame duration $T$ and bandwidth $B$.}}} 
{ From (\ref{eqn46})-(\ref{eqn52}), it follows that the OTFS input-output relation can be expressed as a discrete twisted convolution} 
\begin{eqnarray}
\label{eqn53}
\hspace{-6mm}
y_{_{\mbox{\footnotesize{dd}}}}[k',l'] & \hspace{-3mm} = & \hspace{-3mm} \sum\limits_{k, l \in {\mathbb Z}}  h_{_{\mbox{\footnotesize{dd}}}}[k, l] \, x_{_{\mbox{\footnotesize{dd}}}}[k' - k ,l' - l]
\, e^{j 2 \pi \frac{(k' - k)}{M} \frac{l}{N} } \nonumber \\
& \hspace{-3mm} = & \hspace{-3mm} \sum\limits_{k, l \in {\mathbb Z}}  h_{_{\mbox{\footnotesize{dd}}}}[k' - k, l' - l] \, x_{_{\mbox{\footnotesize{dd}}}}[k ,l]
\, e^{j 2 \pi \frac{(l' - l)}{N} \frac{k}{M}},  
\end{eqnarray}
{where $h_{_{\mbox{\footnotesize{dd}}}}[k, l]$ is the discrete effective DD domain channel filter, given by sampling the continuous effective DD domain channel filter $h_{_{\mbox{\footnotesize{dd}}}}(\tau, \nu)$, i.e.,}
\begin{eqnarray}
\label{eqn53_b}
h_{_{\mbox{\footnotesize{dd}}}}[k, l]  & \Define &  h_{_{\mbox{\footnotesize{dd}}}}(\tau,\nu){\Big \vert}_{\left( \tau = \frac{k \tau_p}{M} \,,\, \nu = \frac{l \nu_p}{N} \right)}\,,\, \mbox{\small} \nonumber \\
h_{_{\mbox{\footnotesize{dd}}}}(\tau , \nu) & \Define &  w_{rx}(\tau, \nu) *_{\sigma} h(\tau,\nu) *_{\sigma}  w_{tx}(\tau, \nu).
\end{eqnarray}
{Typically, the transmit and receive filters are localized and the channel admits bounded delay and Doppler spreads, and hence, in this case, $h_{_{\mbox{\footnotesize{dd}}}}(\tau , \nu)$ is zero whenever $\tau > \tau_{max}, \tau < 0$ or $\vert \nu \vert > \nu_{max}$. We refer to $\tau_{max}$ and $2\nu_{max}$ as the effective delay and Doppler spreads, respectively. Consequently, if the periods satisfy the crystallization condition (\ref{eqn28}) with respect to the effective spreads, that is, $\tau_p > \tau_{max}$ and $\nu_p > 2\nu_{max}$, the input-output relation becomes non-fading and predictable  (cf. Section \ref{secChannelInteractionDD}).}
\begin{figure*}[!h]
\centering
\includegraphics[width=15cm, height=9.0cm]{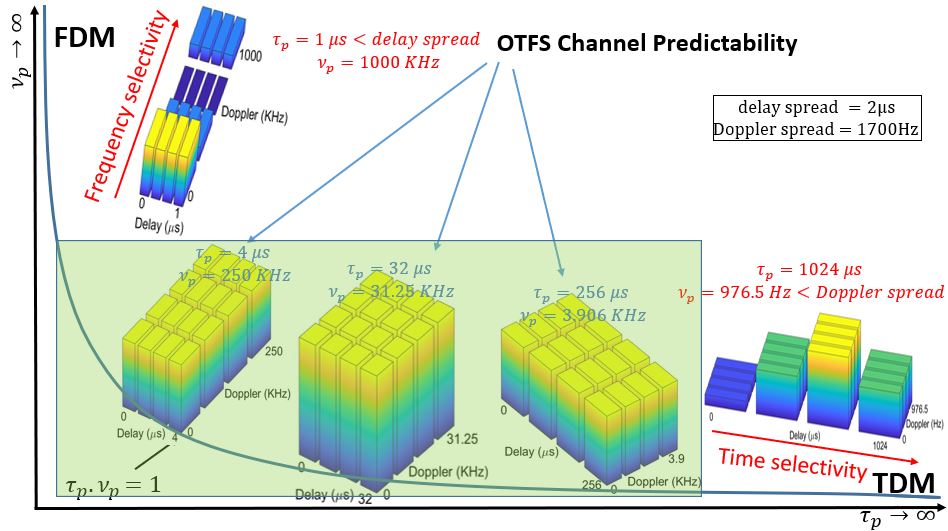}
\caption{For a fixed
$M = N = 4$, the average power of the 
received discrete DD domain signal is plotted for varying delay and Doppler domain period $(\tau_p,\nu_p)$
under the constraint $\tau_p \, \nu_p = 1$.
As $\tau_p \rightarrow 0$, the delay period becomes smaller than the effective channel delay spread due to which
there is aliasing along the delay domain resulting in variation in the received power i.e., fading along the Doppler domain.
We know that OTFS $\rightarrow$ FDM as $\tau_p \rightarrow 0$, which explains frequency selectivity in FDM.
On the other hand, as $\tau_p \rightarrow \infty$, $\nu_p = (1/\tau_p) \rightarrow 0$ and the Doppler period 
becomes smaller than the effective channel Doppler spread
due to which there is aliasing along the Doppler domain resulting in fading along the delay domain. We know that OTFS $\rightarrow$ TDM as $\tau_p \rightarrow \infty$, which explains time selectivity in TDM.
However, when the delay and Doppler periods are greater than the effective channel delay and Doppler spread respectively, there is no aliasing and the OTFS input-output
relation is predictable and exhibits no fading (the average received power profile is flat and appears like the surface of crystalline solids).}
\label{fig_9b}
\end{figure*}

\underline{Crystallization of the OTFS input-output relation}:
{Fig.~\ref{fig_9b} depicts the non-fading and predictability attributes of the OTFS input-output relation for the channel example from Fig.~\ref{fig_7}. Recall, that the delay and Doppler spreads of this channel are $2 \mu s$ and $1700$ Hz, respectively. The main point is that despite this channel being doubly-spread, the OTFS input-output relation is non-fading and predictable whenever the DD periods satisfy the crystallization condition with respect to the channel spreads. For this example, it means that $\tau_p > 2 \mu s$ and $\nu_p > 1700$ Hz. Moreover, the constraint $\tau_p \, \nu_p = 1$ implies that the \emph{crystalline regime} is $2 \mu s < \tau_p < 588 \mu s$ and $1700 < \nu_p < 5 \times 10^5$ Hz (see the green rectangle in Fig.~\ref{fig_9b}). Operating the system in the crystalline regime allows to maintain uniform performance over a wide a range of channel spreads and multitude of use-cases (e.g., Leo-satellites/ UAV communication, mmWave/THz communication). These practical aspects will be discussed in more detail in the second part of this paper.} Table \ref{tab2} summarizes attributes of the input-output relations for TDM, FDM and OTFS. 

\begin{table}
\caption{Attributes of input-output relation.}
\centering
\begin{tabular}{ | c || c | c |  c |  c | } 
  \hline
   Attribute & Channel type & TDM & FDM & OTFS  \\
   \hline
   \hline
  \multirow{3}{5em}{Non-fading} & Delay spread only &  $\checkmark$  &  $\times$   &  $\checkmark$   \\
  \cline{2-5} 
                             & Doppler spread only &  $\times$   &   $\checkmark$    &    $\checkmark$  \\
                             \cline{2-5} 
                             & Doubly-spread  &  $\times$   &  $\times$   &   $\checkmark$  \\
\hline
  \multirow{3}{5em}{Predictable} & Delay spread only &  $\checkmark$  &  $\times$   &  $\checkmark$   \\
  \cline{2-5} 
                             & Doppler spread only &  $\times$   &   $\checkmark$    &    $\checkmark$  \\
                             \cline{2-5} 
                             & Doubly-spread  &  $\times$   &  $\times$   &   $\checkmark$  \\
\hline
\end{tabular}
\label{tab2}
\end{table}

\section{Conclusions}
{\color{black} 
The roots of OTFS modulation go back to the Erlangen Program, introduced by Felix Klein in 1872, which sought to understand mathematical structures, like channels, through the symmetries which leave them invariant. It provides an umbrella for several results in information theory and coding, beginning with the fact that it is Gaussian inputs that achieve capacity on Gaussian channels \cite{Shannon}. It suggests that we should take advantage, when channel errors form a group. Quantum computing provides an example, since bit flips provide discrete analogs of delay operators, and phase flips provide discrete analogs of Doppler operators. A commutative subgroup of the Pauli group determines a quantum error correcting codes, which stores information on common eigenmodes of the subgroup \cite{CSS}. The parallels with OTFS are very clear.}

{ We have described the OTFS modulation within a mathematical framework of Zak theory. Within this framework, the OTFS carrier is a quasi-periodic DD domain pulse which when converted to time via the inverse Zak transform is realized by a pulsone. The main technical message of this paper is that whenever the DD periods of the pulse are taken to be large compared with the channel spreads, the OTFS input-output relation is non-fading and predictable. When this constraint holds, we say that one operates in the crystalline regime. The follow-up of this paper will demonstrate in detail the performance advantages of operating in the crystalline regime.}

{ Compatibility with contemporary multi-carrier signaling motivated an approximation \cite{RH1}, which we refer to as MC-OTFS, to the Zak theoretic variant of OTFS described in this paper, which we refer to as Zak-OTFS. In MC-OTFS, DD domain signals are periodic functions of two variables (instead of quasi-periodic functions) and the conversion to the TD is carried in two steps (instead of one step Zak transform): the first step is conversion to the TF domain using the inverse symplectic finite Fourier transform, and the second step is conversion from TF domain to TD using the Heisenberg transform. In the follow-up of this paper, we will provide a detailed performance comparison between Zak-OTFS and MC-OTFS variants. 

The first wave of OTFS research mainly focused on MC-OTFS. We expect the next wave of OTFS research to be focused on Zak-OTFS, which can offer performance and complexity advantages.}

\section{Acknowledgements}
The first author would like to thank Mr. Imran Ali Khan (Research Scholar at I.I.T. Delhi) for his help in making some of the figures in this paper. The work of Saif Khan Mohammed was supported by the Prof. Kishan and Pramila Gupta Chair at I.I.T. Delhi. The second author, Ronny Hadani, would like to thank Shlomo Rakib and Shachar Kons from Cohere Technologies as many of the concepts described in this paper are inspired from several years of development conducted in collaboration with them. A. Chockalingam acknowledges the support from the J. C. Bose National Fellowship, Science and Engineering Research Board, Department of Science and Technology, Government of India. The work of Robert Calderbank was supported in part by the Air Force Office of Scientific Research under grants FA 8750-20-2-0504 and FA 9550-20-1-0266.

\begin{IEEEbiography}[{\includegraphics[width=1.0in,height=1.25in]{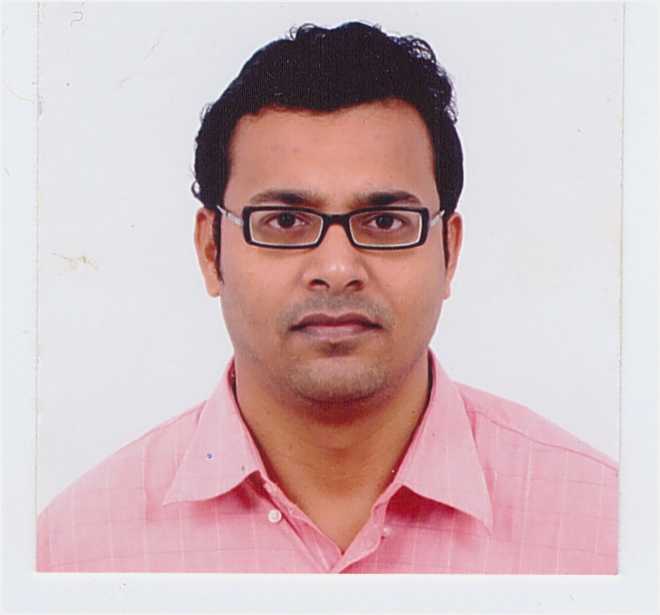}}]{Saif Khan Mohammed}
is a Professor with the
Department of Electrical Engineering, Indian Institute of Technology Delhi (IIT Delhi). He currently holds the Prof. Kishan and Pramila Gupta Chair
at IIT Delhi.
He received the B.Tech. degree in Computer Science and Engineering from IIT Delhi,
New Delhi, India, in 1998, and the Ph.D. degree from the Electrical and Communication
Engineering Department, Indian Institute of Science, Bangalore, India, in 2010.
From 2010 to 2011, he was a
Post-Doctoral Researcher at the Communication Systems Division (Commsys), Electrical Engineering
Department (ISY), Linkoping University, Sweden. He was an Assistant Professor at Commsys, from
September 2011 to February 2013. His main research interests include, waveforms for high
mobility scenarios in sixth generation (6G) communication systems, wireless communication
using large antenna arrays, coding and signal processing for wireless communication
systems, information theory, and statistical signal processing. He currently serves as an
Editor for IEEE Transactions on Wireless Communications and in the past he has served as an Editor for IEEE
Wireless Communications Letters and Physical Communication journal (Elsevier). He holds four
granted U.S. patents in multi-user detection and precoding for multiple-input multiple-output
(MIMO) communication systems. He received the 2017 NASI Scopus Young Scientist Award and the
Teaching Excellence Award at IIT Delhi for the year 2016–2017. He was also a recipient of the
Visvesvaraya Young Faculty Fellowship from the Ministry of Electronics and IT, Government of India,
from 2016 to 2019. Contact him at saifkmohammed@gmail.com.
\end{IEEEbiography}

\begin{IEEEbiography}[{\includegraphics[width=1in,height=1.25in]{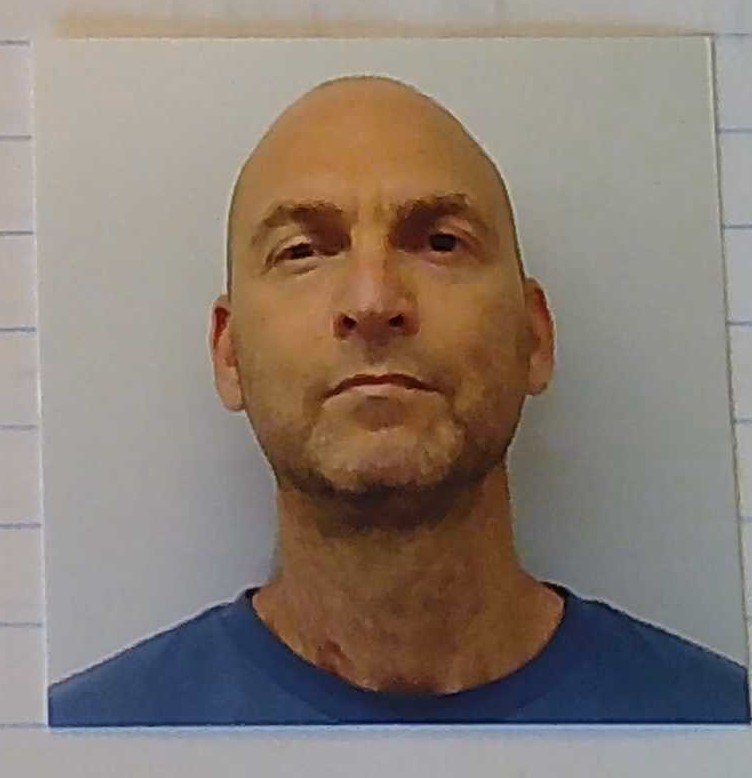}}]{Ronny Hadani}
is an associate professor in the Mathematics Department of the University of Texas at Austin. He also serves as the Chief Science Officer at Cohere Technologies. He holds a PhD in pure mathematics from Tel-Aviv University and a Master degree in applied mathematics from The Weizmann Institute of Science. Contact him at hadani@math.utexas.edu. 
\end{IEEEbiography}

\begin{IEEEbiography}[{\includegraphics[width=1in,height=1.25in]{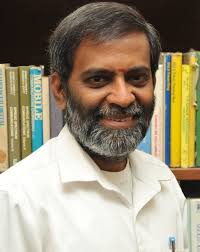}}]{Ananthanarayanan Chockalingam}
was born in Rajapalayam, Tamil Nadu, India. He received the B.E. (Honors) degree in electronics and communication engineering from the P. S. G. College of Technology, Coimbatore, India, in 1984, the M. Tech. degree in electronics and electrical communications engineering (with specialization in satellite communications) from the Indian Institute of Technology, Kharagpur, India, in 1985, and the Ph.D. degree in electrical communication engineering (ECE) from the Indian Institute of Science (IISc), Bangalore, India, in 1993. During 1986 to 1993, he worked with the transmission R\&D division of the Indian Telephone Industries Limited, Bangalore. From December 1993 to May 1996, he was a Postdoctoral Fellow and an Assistant Project Scientist in the department of electrical and computer engineering, University of California, San Diego. From May 1996 to December 1998, he served Qualcomm, Inc., San Diego, CA, as a Staff Engineer/Manager in the systems engineering group. In December 1998, he joined the faculty of the department of ECE, IISc, Bangalore, India, where he is a Professor, working in the area of wireless communications.

Dr. Chockalingam is a recipient of the Swarnajayanti Fellowship from the Department of Science and Technology, Government of India. He served as an Associate Editor of the IEEE TRANSACTIONS ON VEHICULAR TECHNOLOGY, and as an Editor of the IEEE TRANSACTIONS ON WIRELESS COMMUNICATIONS. He served as a Guest Editor for the IEEE JOURNAL ON SELECTED AREAS IN COMMUNICATIONS (Special Issue on Multiuser Detection for Advanced Communication Systems and Networks), and for the IEEE JOURNAL OF SELECTED TOPICS IN SIGNAL PROCESSING (Special Issue on Soft
Detection on Wireless Transmission). He is a Fellow of the Indian National Academy of Engineering, the National Academy of Sciences, India, the Indian National Science Academy, and the Indian Academy of Sciences. Contact him at achockal@iisc.ac.in.
\end{IEEEbiography}

\begin{IEEEbiography}[{\includegraphics[width=1in,height=1.25in]{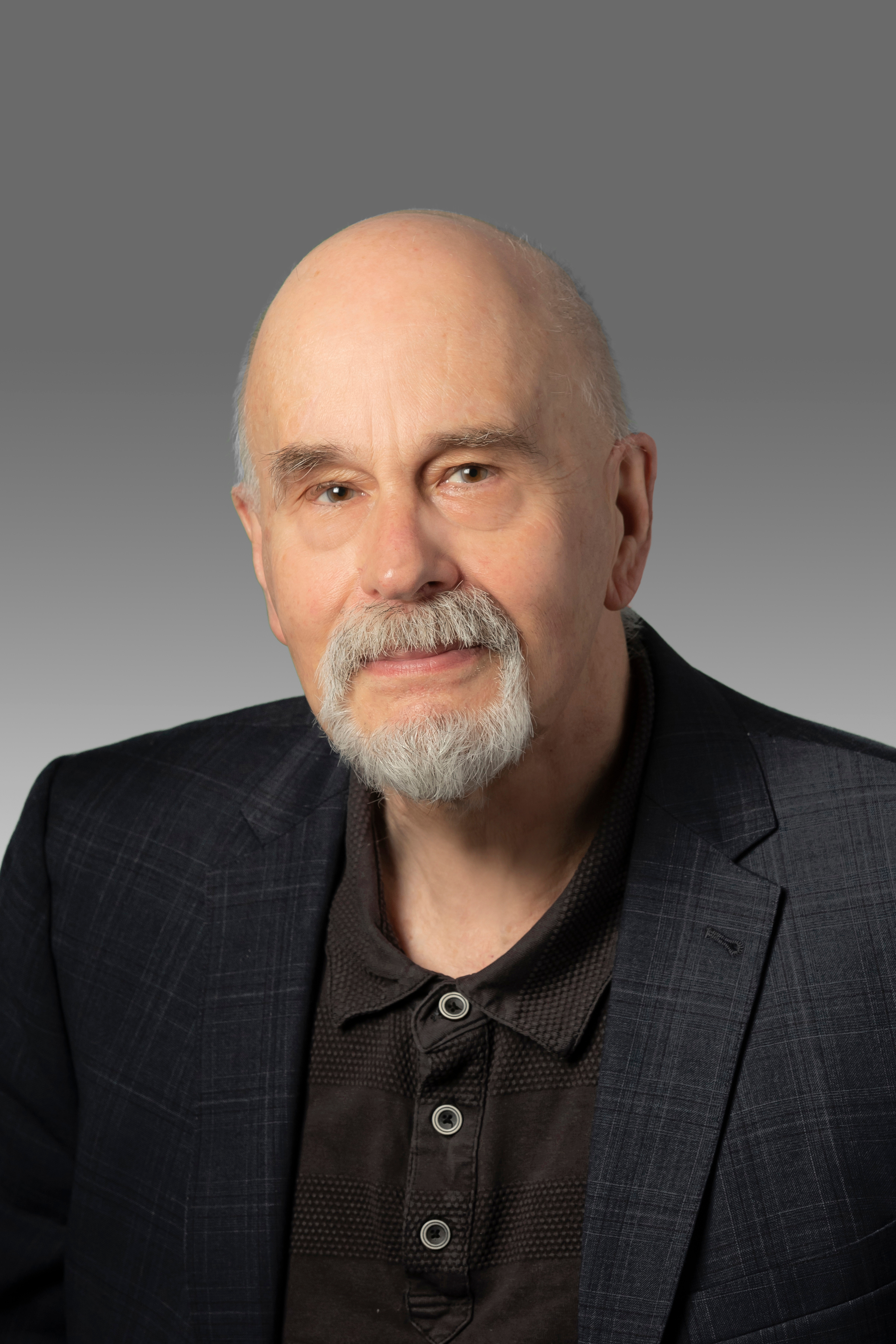}}]{Robert Calderbank}
directs the Rhodes Information initiative at Duke University, where he is a Distinguished Professor of Electrical and Computer Engineering, Computer Science and Mathematics.
 
He started his career in the Mathematical Sciences Research Center at Bell Labs, and he left AT\&T in 2003 as Vice President for Research. Dr. Calderbank directed the Program in Applied and Computational Mathematics at Princeton University before joining Duke University in 2010.  He was elected to the National Academy of Engineering in 2005, and to the American Academy of Arts and Sciences in 2022. Dr. Calderbank received the 2015 Hamming Medal, and the 2015 Shannon Award.
 
Dr. Calderbank is known for contributions to voiceband modem technology at the dawn of the internet, and for contributions to wireless communication that are incorporated in billions of cell phones. He has also made contributions to quantum error correction that provide a foundation for fault tolerant quantum computation. Contact him at robert.calderbank@duke.edu.
\end{IEEEbiography}

\end{document}